\documentclass[journal,twocolumn]{IEEEtran}

\usepackage[colorlinks,urlcolor=blue,linkcolor=blue,citecolor=blue]{hyperref}
\usepackage{soul}
\usepackage{subcaption}
\usepackage{graphicx}
\usepackage{hyperref}
\usepackage{amsmath, amssymb,amsthm,cite}
\usepackage{amsmath, amssymb,amsthm,cite}
\usepackage[utf8x]{inputenc}
\usepackage{placeins} 
\usepackage{graphicx}
\usepackage{psfrag}
\usepackage{bbm}
\usepackage{amsmath,amsfonts,amsthm,bm}
\usepackage{lipsum}
\usepackage{dblfloatfix}
\usepackage{mathtools}
\usepackage{algpseudocode}
\usepackage{algorithm}
\usepackage{color}
\usepackage{tcolorbox}
\usepackage{etoolbox}
\usepackage{physics}
\usepackage{tabularx}
\usepackage{multirow}
\usepackage{mleftright}

\allowdisplaybreaks
\newcommand{\RNum}[1]{\uppercase\expandafter{\romannumeral #1\relax}}
\usepackage{hyperref}
\usepackage{authblk}
\usepackage[english]{babel}
\usepackage{graphicx}
\usepackage{academicons}
 \usepackage{setspace}
\definecolor{orcidlogocol}{HTML}{A6CE39}

\begin{document}
\title{Neural Network-Based Multi-Target Detection within Correlated Heavy-Tailed Clutter}

\author{S. Feintuch\thanks{Stefan Feintuch, Haim H. Permuter, Igal Bilik, and Joseph Tabrikian are with the School of Electrical and Computer Engineering, Ben Gurion University of the Negev, Beer Sheva, Israel. (e-mails: stefanfe@post.bgu.ac.il, haimp@bgu.ac.il, bilik@bgu.ac.il, joseph@bgu.ac.il). This work was partially supported by the Israel Science Foundation under Grants 2666/19 and 1895/21.},
H. Permuter, {\it Senior Member, IEEE}, I. Bilik, {\it Senior Member, IEEE}, and J. Tabrikian, {\it Fellow, IEEE}
}

\maketitle

\begin{abstract}This work addresses the problem of range-Doppler multiple target detection in a radar system in the presence of \textit{slow-time} correlated and heavy-tailed distributed clutter.
Conventional target detection algorithms assume Gaussian-distributed clutter, but their performance is significantly degraded in the presence of correlated heavy-tailed distributed clutter.
Derivation of optimal detection algorithms with heavy-tailed distributed clutter is analytically intractable. Furthermore, the clutter distribution is frequently unknown.
\textcolor{black}{This work proposes a deep learning-based approach for multiple target detection in the range-Doppler domain. The proposed approach is based on a unified NN model to process the time-domain radar signal for a variety of signal-to-clutter-plus-noise ratios (SCNRs) and clutter distributions, simplifying the detector architecture and the neural network training procedure.} \textcolor{black}{The performance of the proposed approach is evaluated in various experiments using recorded radar echoes, and via simulations, it is shown that the proposed method outperforms the conventional cell-averaging constant false-alarm rate (CA-CFAR), the \textcolor{black}{trimmed-mean} CFAR (\textcolor{black}{TM-CFAR}), and the adaptive normalized matched-filter (ANMF) detectors in terms of probability of detection in the majority of tested SCNRs and clutter scenarios.}
\end{abstract}


\section{INTRODUCTION}\label{sec:intro}

Target detection in range-Doppler map is one of the major radar tasks~\cite{van2004detection,Skolnik:98948}.
Conventionally, the decision on target presence is made by comparing the energy within the cell-under-test with a threshold, which is calculated according to the energy at neighboring cells~\cite{richards2010principles}.
The presence of spiky clutter in the cells used for the detection threshold calculation increases the threshold level, and thus, compromises the target detection performance~\cite{richards2010principles}.


Considering the detector input as a one-dimensional complex signal that contains \textit{slow-time} samples of received radar echoes in each range bin, the task of radar target detection within correlated heavy-tailed clutter is conventionally formulated as a binary hypotheses decision task. Under this formulation, the hypotheses $\mathcal{H}_0$ and $\mathcal{H}_1$ represent cases where there is no target and the target is present within the cell-under-test (CUT), respectively~\cite{richards2010principles,conte1995asymptotically, sangston2012coherent, sangston2015adaptive, kraut1999cfar, conte1998adaptive, conte2002recursive,gini2002covariance,kammoun2017optimal,coluccia2021knn, 9054283, coluccia2020k, coluccia2019radar,petrov2017detection,gandhi1997neural, de2005cfar,guo2020anomaly,ahmed2021reinforcement,fortunati2020massive,8506468}.
In~\cite{conte1995asymptotically, sangston2012coherent, sangston2015adaptive, kraut1999cfar, conte1998adaptive, conte2002recursive,gini2002covariance} the problem of radar target detection was formulated as a binary hypothesis testing, where the optimum detectors were derived under certain conditions.
\textcolor{black}{The design for a regularized covariance matrix estimation in the adaptive normalized matched-filter (ANMF) was introduced in~\cite{kammoun2017optimal} to maximize the asymptotic probability of detection, while retaining a constant false-alarm rate (CFAR).}
The properties of CFAR detectors in the presence of correlated heavy-tailed clutter were studied in~\cite{de2005cfar}. 
The problem of range-migrating target detection within heavy-tailed clutter was addressed in~\cite{petrov2017detection}, in which a fast-converging amplitude estimation algorithm for target detection was proposed.
An orthogonal-projection-based approach to suppress the sea clutter at each range cell in combination with cell-averaging CFAR (CA-CFAR), was proposed in~\cite{8506468}.
The authors in \cite{fortunati2020massive} addressed the target detection within heavy-tailed clutter using massive multiple-input multiple-output (mMIMO) radar. Therein, a detector was proposed for the asymptotic regime with increasing number of antennas, and its robustness to the unknown clutter distribution was demonstrated.

However, these model-based approaches were designed considering a specific measurement model, and their performance may degrade in the case of model mismatch.
Alternatively, data-driven machine learning (ML) approaches have been proposed in~\cite{coluccia2021knn, 9054283, coluccia2020k, coluccia2019radar,guo2020anomaly,wang2017intelligent,ahmed2021reinforcement}.
In these approaches, target detection is performed using features extracted from the data. Thus, they enable detectors' robustness to environmental and clutter statstics' variations.
{\it K}-nearest neighbors (KNN) based approaches using various feature space transforms of the raw one-dimensional complex signal were proposed in~\cite{coluccia2021knn, 9054283, coluccia2020k, coluccia2019radar,guo2020anomaly} to address the binary hypothesis decision task. 
In particular, the authors in~\cite{coluccia2021knn, coluccia2020k} proposed to obtain a KNN-based decision rule from simulated data, and evaluated the proposed methods using the IPIX database \cite{IPIX} of recorded radar echoes that contain correlated heavy-tailed sea clutter.
Authors in~\cite{wang2017intelligent} used support vector machine to switch between conventional CFAR methods and perform target detection in an environment containing clutter edges and/or multiple interfering targets under white Gaussian noise.
The work in~\cite{ahmed2021reinforcement} extended the work in~\cite{fortunati2020massive} to angle dimension and proposed a reinforcement learning (RL) based approach to design the beamforming matrix in a cognitive radar (CR) setup.

\textcolor{black}{The binary hypothesis-based approaches in~\cite{conte1995asymptotically, sangston2012coherent, sangston2015adaptive, kraut1999cfar, conte1998adaptive, conte2002recursive,gini2002covariance,kammoun2017optimal,coluccia2021knn, 9054283, coluccia2020k, coluccia2019radar,petrov2017detection,gandhi1997neural, de2005cfar,ahmed2021reinforcement,8506468,guo2020anomaly,fortunati2020massive} 
assume under the $\mathcal{H}_1$ hypothesis a) the presence of a single-target within each CUT and b) the availability of target-free secondary data, which is used for clutter covariance matrix estimation. However, practical scenarios may include multiple targets with similar azimuth, range, and Doppler. Therefore, the performances of these methods degrade in such scenarios.}
In addition, the methods in \cite{conte1995asymptotically, sangston2012coherent, sangston2015adaptive, kraut1999cfar, conte1998adaptive, conte2002recursive,gini2002covariance,coluccia2021knn, 9054283, coluccia2020k, coluccia2019radar,petrov2017detection,gandhi1997neural, de2005cfar,guo2020anomaly,8506468,wang2017intelligent} use the data after range matched-filter processing, which linearly projects each \textit{fast-time} received pulse to range bins \cite{richards2010principles}. This linear transformation fails to suppress the clutter echo signals, since these are not orthogonal to the projection signals that correspond to each range bin.

Recently, deep neural networks (DNNs) with various network architectures have been introduced for radar target detection, where the network input consists of the samples of the received radar echo~\cite{gandhi1997neural,cao2021dnn,jiang2021method,ofek2011modular}.
Considering a one-dimensional problem with \textit{a-priori} known signal, a multi-layer perceptron (MLP) based detector for binary hypothesis detection within non-Gaussian noise was proposed in~\cite{gandhi1997neural}. 
A fully-connected architecture for multiple target detection in the presence of homogeneous Rayleigh-distributed clutter was utilized in~\cite{cao2021dnn}. A single-target detection within additive white Gaussian noise (AWGN) using convolutional neural network (CNN) based architecture for range-Doppler target detection and azimuth-elevation estimation was proposed in~\cite{jiang2021method}. 
However, the works in \cite{cao2021dnn,jiang2021method,ofek2011modular} assume white Gaussian-distributed clutter, whereas a more realistic clutter model would be correlated and non-Gaussian. Although the work in~\cite{gandhi1997neural} addresses the non-Gaussian clutter, it also assumes the binary hypothesis decision task, which has limitations as previously mentioned.


Neural network (NN) based processing using range-Doppler map input was also studied in the literature, the majority of these works invoke computer vision methods for radar target detection within AWGN~\cite{wang2019study,8891420,brodeski2019deep,9207080}.
A fully connected NN architecture for multiple target detection within heavy-tailed clutter was proposed in~\cite{akhtar2019go}.
A residual block~\cite{he2016deep} was proposed in~\cite{8891420} for background noise estimation in the range-Doppler map for the conventional CFAR detector.
In~\cite{brodeski2019deep} a model-based data augmentation technique was proposed for linear frequency modulated (LFM) radar detector in the 3D range-Doppler-angle domain. The proposed technique was used to generate a synthetic dataset for U-net~\cite{ronneberger2015u} training, considering a single target in the azimuth-elevation domain at each range-Doppler region-of-interest (ROI).
The work in~\cite{9207080} extended~\cite{brodeski2019deep}, by utilizing the absolute value of the range-Doppler map for additional data augmentation.

Contrary to previous works described above, which address the radar target detection within heavy-tailed clutter as a one-dimensional binary hypothesis decision task for each range bin, this work addresses the problem of radar target detection within heavy-tailed clutter as a two-dimensional (range-Doppler) detection problem with multiple targets in unknown ranges and radial velocities (Doppler).
Furthermore, in practical radar scenarios characterized by correlated clutter, the conventional range-Doppler transform designed for AWGN model fails to suppress the clutter, since the clutter signal is correlated in \textit{slow-time} and can be spread over multiple range bins. 
Therefore, the range-Doppler map-based DNN approaches mentioned above, do not fully exploit the power of DNNs to learn highly abstract nonlinear transformations for suppressing the clutter.
To that aim, this work proposes to leverage DNN's ability to learn highly complex nonlinear functions in order to transform the complex time-domain radar echo samples into the range-Doppler domain while suppressing the correlated clutter.

The contributions of this work are:
\begin{enumerate}
    \item A novel neural processing block named dimensional-alternating fully connected (DAFC) block, is proposed to process raw time-domain radar echoes for the task of multiple target detection. A DNN architecture that utilizes this block is proposed to map radar signals to either range or Doppler domains while suppressing correlated heavy-tailed clutter.
    \item The proposed DNN architecture is utilized as part of a novel range-Doppler multiple target detector, that is evaluated in the presence of correlated heavy-tailed clutter.
    \item The proposed method significantly outperforms conventional methods and proves to be more robust in various aspects: multiple targets within AWGN and correlated heavy-tailed clutter, varying clutter conditions/``spikiness'' measure, and detection threshold sensitivity to clutter ``spikiness''. 
    \item The proposed method proves to generalize well to unseen data, based on experiments involving recorded real data.
\end{enumerate}


The following notations will be used throughout the paper. Roman boldface  lower-case and upper-case letters represent vectors and matrices, respectively.  Non-bold italic letters stands for scalars. 
$\mathbf{I}_N$ and $\mathbf{0}_N$ are the identity matrix and zero matrix of size $N \times N$, respectively.
$\mathbb{E}$, superscript $T$, and superscript $H$ are the expectation, transpose, and Hermitian transpose operators, respectively.
$\text{Vec}$, $|\cdot|$, and $\mathbb{I}$ stand for the vectorization, set size, and indicator operators, respectively.
$\left(\cdot\right)$ and $\left(\cdot,\cdot\right)$ denote single argument and double arguments functions.
$[\mathbf{a}]_{n}$ and $[\mathbf{A}]_{n,m}$ are the $n$-th and $n,m$-th elements of the vector $\mathbf{a}$ and the matrix $\mathbf{A}$, respectively.
$[\mathbf{A}]_{[\cdot,:]}$ and $[\mathbf{A}]_{[:,\cdot]}$ represent an arbitrary row and column in the matrix $\mathbf{A}$, respectively.

The remainder of this paper is organized as follows: The addressed problem is stated in Section \ref{sec:prob_def}. Section \ref{sec:novel_nn_design} presents the proposed DAFC-based radar target detection approach. The performance of the proposed approach is evaluated via simulated data and recorded real data in Section \ref{sec:eval}, and our conclusions are summarized in Section \ref{sec:conclusion}.

\section{PROBLEM STATEMENT}\label{sec:prob_def}
The measurement model is described in Subsection \ref{subsec:meas_model}, and the multiple target detection problem in the range-Doppler domain is formulated in Subsection \ref{subsec:rd_detection_formulation}.

\begin{figure}[!t]
\centering
    \includegraphics[width=0.45\textwidth]{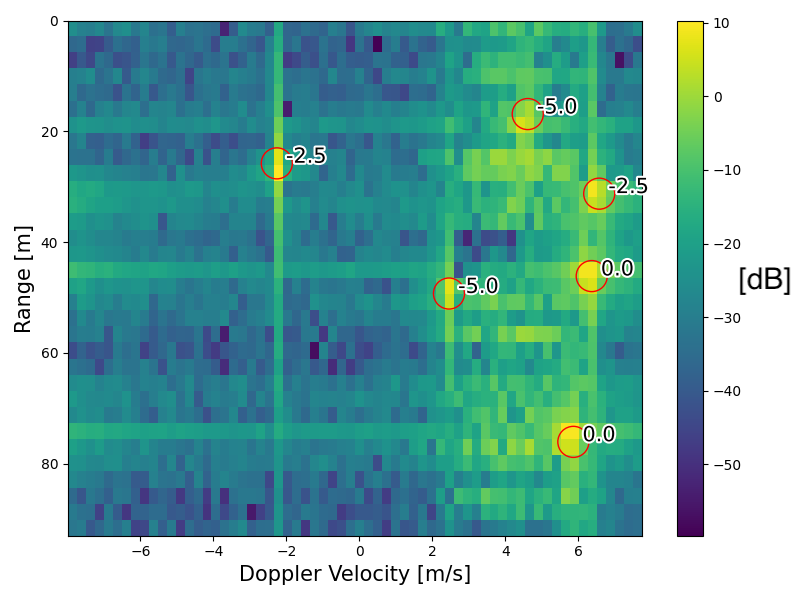}
    \caption{Example of range-Doppler map containing simulated targets, clutter, and noise. The targets are circled in red, with the marked $\text{SCNR}$s.
    \textcolor{black}{The non-homogeneous clutter is present at the vicinity of the $4m/s$ Doppler velocity in the majority of the range bins.}} \label{fig:RD_map_with_target}
\end{figure}

\begin{figure*}[!t]
\centering
\includegraphics[width=0.7\textwidth]{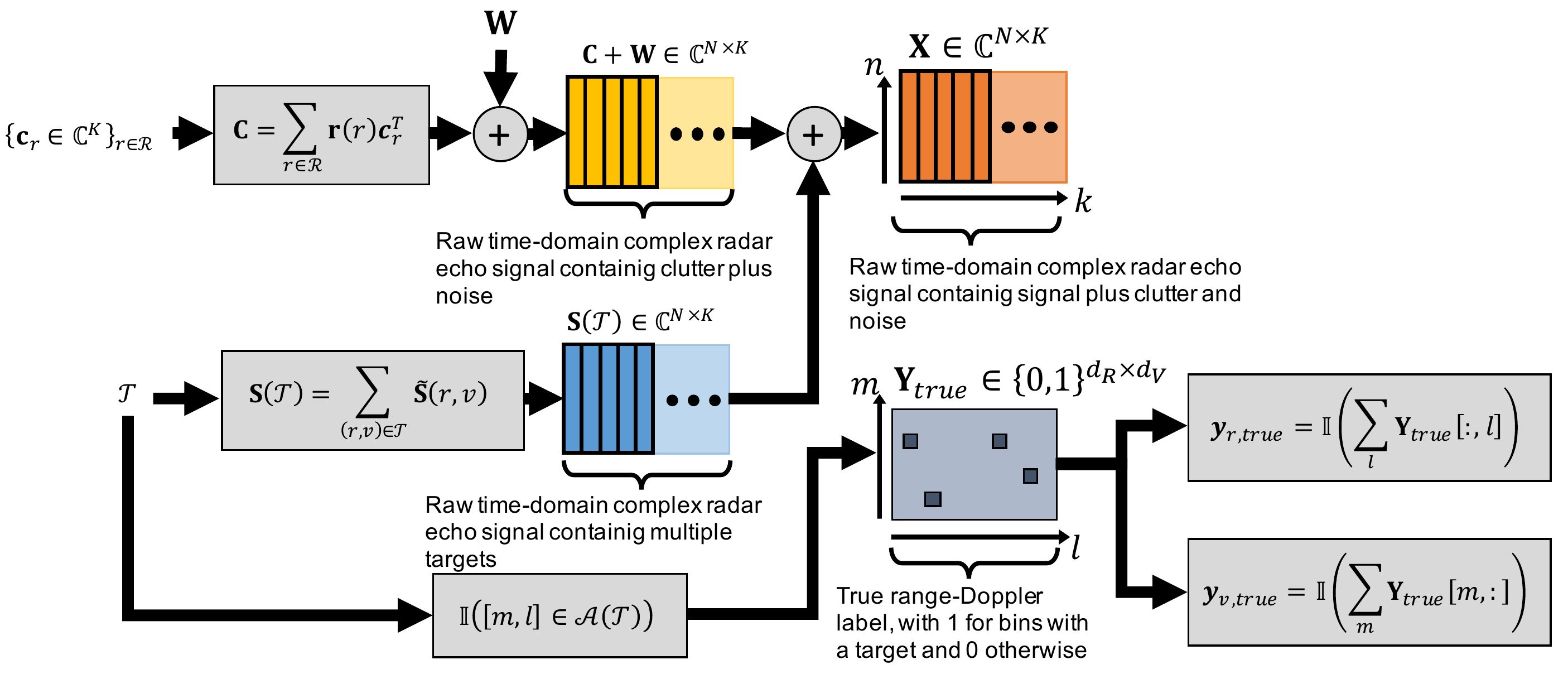}
\caption{Dataset generation scheme.}
\label{fig:dataset_gen}
\end{figure*}

\subsection{Measurement Model}\label{subsec:meas_model}
\textcolor{black}{Consider the baseband \textit{fast-time} $\times$ \textit{slow-time} model of a single received radar echo:}
\begin{align}\label{eq:meas_model_SCW}
    &\mathbf{X} = \mathbf{S}\left(\mathcal{T}\right) + \mathbf{C} + \mathbf{W}\;
\end{align}
where $\mathbf{X},\mathbf{S}\left(\mathcal{T}\right),\mathbf{C},\mathbf{W}\in\mathbb{C}^{N\times K}$, $\mathcal{T} = \left\{ (r_j,v_j) : (r_j,v_j)\in \left[r_{min}, r_{max}\right]\times \left[v_{min}, v_{max}\right] \right\}$ denotes the set of targets present in the frame, and $\left[r_{min}, r_{max}\right]$ and $\left[v_{min}, v_{max}\right]$ are intervals of targets' ranges and radial velocities, respectively. 
The matrices $\mathbf{S}\left(\cdot\right)$, $\mathbf{C}$, and $\mathbf{W}$ represent the target echo signal, the clutter, and the additive noise.
The targets' matrix $\mathbf{S}\left(\cdot\right)$ is defined as:
\begin{align}\label{eq:meas_model_S}
    &\mathbf{S}\left(\mathcal{T}\right) = 
    \begin{cases}
        \sum_{(r,v)\in\mathcal{T}}{\widetilde{\mathbf{S}}(r,v)}\;&, \mathcal{T}\ne\emptyset\\
        \mathbf{0}_{N\times K} &,\mathcal{T}=\emptyset
    \end{cases}
\end{align}
where $\widetilde{\mathbf{S}}(r,v)$ is the radar echo matrix received from a single target at range $r$ and radial velocity $v$, and is defined as \cite{richards2010principles}:
\begin{align}\label{eq:target_echo_signal}
    \widetilde{\mathbf{S}}(r,v) =&
    A_{rv}e^{j\phi_{rv}} \mathbf{r}(r)\mathbf{v}^T(v)
\end{align}
where $\mathbf{0}_{N\times K}$ dentoes the $N \times K$ zero matrix, $\phi_{rv}\sim\text{U}\left([0,2\pi]\right)$ is unknown phase, $A_{rv}\in\mathbb{R}^+$ represents the received signal amplitude and depends on the target radar cross section (RCS) and the propagation path loss. 

\textcolor{black}{Notice that the model in~\eqref{eq:meas_model_SCW} represents the radar echo of both the pulse-Doppler and the LFM-CW radars with the appropriate range and radial velocity steering vectors, $\mathbf{r}(\cdot)$ and $\mathbf{v}(\cdot)$. Thus, for LFM-CW radar}:
\begin{align}\label{eq:target_echo_signal_parameters}
    & \mathbf{r}(r) = \left[1\quad e^{-j2\pi\frac{2Br}{cN}}\quad\dots\quad e^{-j2\pi\frac{2Br}{cN}(N-1)}\right]^T\;,\\
    & \mathbf{v}(v) = \left[1\quad e^{-j2\pi\frac{2f_cv}{c}T_0}\quad\dots\quad e^{-j2\pi\frac{2f_cv}{c}T_0(K-1)} \right]^T\;,\nonumber
\end{align}
where $N$ is the number of samples per \textcolor{black}{LFM chirp}, $K$ is the number of chirps per dwell, $B$ is the transmit signal bandwidth, $f_c$ is the carrier frequency, $c$ is the speed of light, and $T_0$ stands for the pulse repetition interval (PRI). 

Conventionally, \textit{slow-time} radar clutter is statistically modeled as a random vector at each range bin~\cite{coluccia2021knn,conte1991modelling,IPIX}.
Let $\{\mathbf{c}_r\in\mathbb{C}^K\}_{r\in\mathcal{R}}$ denote the group of one-dimensional \textit{slow-time} clutter vectors. Then, the clutter matrix $\mathbf{C}$ in \eqref{eq:meas_model_SCW} can be obtained by converting $\{\mathbf{c}_r\}_{r}$ to the \textit{fast-time} $\times$ \textit{slow-time} representation by:
\begin{align}
    \mathbf{C} = \sum_{r\in \mathcal{R}}{\mathbf{r}\left(r\right)\mathbf{c}_r^T}\;,\label{Clutter}
\end{align}
where $\mathcal{R}$ is the set of range bins, that partition the continuous range space to grid points spaced by the range resolution $\Delta r = c/(2B)$.
The clutter signal matrix $\mathbf{C}$ in \eqref{eq:meas_model_SCW} is a sum of $|\mathcal{R}|$ ``clutter echoes", one per range bin. 
\textcolor{black}{
According to \eqref{Clutter}, each column in $\mathbf{C}$ is a linear combination of the range steering vectors corresponding to the range bins in $\mathcal{R}$. Therefore}, by projecting the \textit{fast-time} vectors (i.e. columns) in~\eqref{Clutter} to the range steering vectors representing the range bins in $\mathcal{R}$, we will obtain the set of original clutter vectors $\{\mathbf{c}_r\}_r$, one per range-bin.



The \textit{fast-time}$\times$\textit{slow-time} noise matrix $\mathbf{W}$ in \eqref{eq:meas_model_SCW} is defined by $\widetilde{\mathbf{w}} = \text{Vec}\left(\mathbf{W}\right)$, where $\widetilde{\mathbf{w}}$ is modeled as an AWGN vector:
\begin{align}
    &\widetilde{\mathbf{w}} \sim \mathcal{CN}\left(\mathbf{0}_{NK}, \sigma^2\mathbf{I}_{NK}\right)\;.
\end{align}

Let $\tilde{\mathbf{s}}\left(r,v\right) \triangleq \text{Vec}(\widetilde{\mathbf{S}}\left(r,v\right))$ and $\tilde{\mathbf{c}} \triangleq \text{Vec}(\mathbf{C})$ be the vectorizations of a target and clutter matrices in \eqref{eq:target_echo_signal} and \eqref{Clutter}, respectively. 
The clutter-to-noise ratio (CNR) for a given frame and signal-to-clutter-plus-noise ratio (SCNR) for a given target within the frame are defined as:
\begin{align}\label{eq:scnr_def}
    \text{CNR} &= \frac{\mathbb{E}\left[\|\tilde{\mathbf{c}}\|^2\right]}{\mathbb{E}\left[\|\tilde{\mathbf{w}}\|^2\right]}\;,\\
    \text{SCNR} &= \frac{\mathbb{E}\left[\|\tilde{\mathbf{s}}\left(r,v\right)\|^2\right]}{\mathbb{E}\left[\|\tilde{\mathbf{c}} + \tilde{\mathbf{w}}\|^2\right]}\;.\nonumber
\end{align}

\subsection{Range-Doppler Detection Formulation}\label{subsec:rd_detection_formulation}
The sets of range and Doppler bins are denoted by $\mathcal{R}$ and $\mathcal{V}$, respectively. 
The range bins defined earlier and the Doppler bins $\mathcal{V}$ partition the continuous Doppler space to grid points spaced by the Doppler resolution $\Delta v = c / (2 f_c K T_0)$.
The set of range-Doppler bins is obtained by the Cartesian product $\mathcal{R}\times\mathcal{V}$.
A range-Doppler detector can be formulated as a mapping between the received signal in \eqref{eq:meas_model_SCW} to a per-bin decision in the range-Doppler domain:
\begin{align}
    \hat{\mathbf{Y}} = \mathbf{H}\left(\mathbf{X}\right) \in \left\{0,1\right\}^{d_R\times d_V},
\end{align}
where $d_R=|\mathcal{R}|$, $d_V=|\mathcal{V}|$ and $\mathbf{H}(\cdot)$ is a mapping from a \textit{fast-time} $\times$ \textit{slow-time} input frame $\mathbf{X}$ to a range-Doppler decision matrix $\hat{\mathbf{Y}}$.

Let $[m,l]$ denote a coordinate in the discrete range-Doppler space $\mathcal{R}\times\mathcal{V}$. The decision on target presence in the range-Doppler bin corresponding to the coordinate $[m,l]$ is defined using entries in the range-Doppler decision matrix $\hat{\mathbf{Y}}$:
\begin{align}\label{eq:rd_decision}
    &\begin{cases}
        \text{Target},& [\hat{\mathbf{Y}}]_{m,l} = 1\\
        \text{No target},& [\hat{\mathbf{Y}}]_{m,l} = 0
    \end{cases}\;.\nonumber
\end{align}
An optimum detector, maximizes the probability of detection $P_D$ for a fixed probability of false-alarm $P_{FA}$.

The conventional range-Doppler transform, which maps the received signal in~\eqref{eq:meas_model_SCW} to the range-Doppler domain, can be obtained by taking the absolute squared value of the 2D-FFT of $\mathbf{X}$. 
Fig. \ref{fig:RD_map_with_target} shows an example of the conventional range-Doppler transform of simulated radar signal consisting of multiple targets, correlated heavy-tailed clutter, and AWGN. Note that there is a non-homogeneous clutter that is observed around the Doppler velocity of $4$ $m/s$, and is present in the majority of the range bins. This example visually exemplifies the limitations of conventional range-Doppler processing in suppression of correlated clutter. Therefore, suppression of correlated heavy-tailed clutter involves nonlinear range-Doppler transforms, as proposed in this work.

Conventional range-Doppler detectors, such as CA-CFAR and \textcolor{black}{TM-CFAR}, operate on the output of the conventional range-Doppler transform (absolute square of 2D-FFT).
The decision for target presence in each range-Doppler bin is based on calculating an adaptive threshold by utilizing the energy information in the surrounding bins~\cite{richards2010principles}.
Correlated heavy-tailed clutter can induce spikes and/or high levels in these surrounding cells, and thus compromise the range-Doppler detector's performance.

\section{PROPOSED APPROACH}\label{sec:novel_nn_design}
The proposed approach for target detection within heavy-tailed clutter is detailed in this section. This paper proposes a data-driven approach, and for this purpose the dataset generation method is described in Subsection \ref{sec:data_gen}. Next, 
the pre-processing and innovative DAFC block are introduced in subsections \ref{subsec:pre_process_flow} and \ref{subsubsec:dafc_block}. Finally, the NN architecture and the proposed range-Doppler detector are detailed in subsections \ref{subsec:nn_arch_and_training} and \ref{sec:rd_detection}.

\subsection{Dataset Generation}\label{sec:data_gen}
The dataset generation is schematically shown in Fig.~\ref{fig:dataset_gen}. First, the simulated radar echoes of clutter and noise are generated using the clutter signals, $\{\mathbf{c}_r\}_{r\in\mathcal{R}}$, to build the matrix $\mathbf{C}$ in~\eqref{Clutter} along with AWGN, $\mathbf{W}$. Next, the set of targets, $\mathcal{T}$ is used to build the signal matrix, $\mathbf{S}\left(\mathcal{T}\right)$ in (\ref{eq:meas_model_S}). 
The received radar input frame, $\mathbf{X}$, is generated according to~\eqref{eq:meas_model_SCW}.
The range-Doppler label matrix $\mathbf{Y}_{true}\in\{0,1\}^{d_R\times d_V}$ is a binary matrix with the following entries:

\begin{align}
    &[\mathbf{Y}_{true}]_{m,l} = 
    \begin{cases}
        1,& [m,l]\in\mathcal{A}\left(\mathcal{T}\right) \\
        0,& \text{else}
    \end{cases},
\end{align}
where $\mathcal{A}\left(\mathcal{T}\right)$ is the set of matrix indices that represent the closest range-Doppler bins to the true targets in $\mathcal{T}$.

This work proposes to transform an input \textit{fast-time}$\times$\textit{slow-time} frame to range or Doppler domains and detect targets using a NN-based approach.
It is proposed to use two separate and identical NN models for range and Doppler domains. These models are trained according to the supervised learning framework.
The range label used to train the range model (i.e. range NN) is $\mathbf{y}_{r,true} = \mathbb{I}\left(\sum_{l}{[\mathbf{Y}_{true}]_{[:,l]}}\right)$ and the Doppler label used to train the Doppler model (i.e. Doppler NN) is $\mathbf{y}_{v,true} = \mathbb{I}\left(\sum_{m}{[\mathbf{Y}_{true}]_{[m,:]}}\right)$.

The clutter signal matrix $\mathbf{C}$ in~\eqref{eq:meas_model_SCW} consists of $|\mathcal{R}|$ clutter \textit{slow-time} vectors $\{\mathbf{c}_r\}_{r\in\mathcal{R}}$. Both simulated and recorded radar echoes are used for these clutter vectors.

\textbf{1) Simulated clutter}

 This work adopts the commonly used K-distributed spherically-invariant random vector (SIRV) model for correlated heavy-tailed clutter \textit{slow-time} vector at each range bin~\cite{coluccia2021knn,conte1991modelling}:
\begin{align}\label{eq:cg_clutter}
    &\mathbf{c} \sim\mathcal{K}\left(\nu,\mathbf{M}\right),\quad\mathbf{c} = \sqrt{s}\mathbf{z}\in\mathbb{C}^K\\
    &s\sim\Gamma(\nu,\nu),\;\mathbf{z}\sim\mathcal{CN}\left(\mathbf{0},\mathbf{M}\right)\nonumber\\
    &[\mathbf{M}]_{[p, q]}=\exp{-2\pi^2\sigma_f^2(p - q)^2 - j(p - q)f_dT_0}\;,\nonumber
\end{align}
where $\Gamma\left(\nu,\nu\right)$ denotes the Gamma distribution with shape $\nu$ and rate $\nu$, {\color{black} $\sigma_f^2$ is inversely related to the clutter correlation,} and $f_d$ is the clutter's Doppler frequency shift \cite{coluccia2021knn}.
The shape parameter, $\nu$, controls the ``spikiness" measure of the clutter amplitude distribution \cite{coluccia2021knn,conte1991modelling}.
A set of independent, identically-distributed (i.i.d.) simulated clutter vectors for each range bin in $\mathcal{R}$ is generated as: $\{\mathbf{c}_r\}_{r\in\mathcal{R}}\stackrel{i.i.d.}{\sim}\mathcal{K}\left(\nu,\mathbf{M}\right)$.

\textbf{2) Recorded real data clutter}: 

We use the Grimsby IPIX database of recorded echoes from radar clutter~\cite{IPIX}. This database contains high-resolution radar echoes collected using the McMaster IPIX radar in Grimsby on the shore of Lake Ontario. Each file in the database contains 60,000 pulses of radar echoes, recorded during 60 seconds. Each file contains radar echoes from Lake Ontario collected at various dates, hours, azimuths, and range sections. 
A set $\{\mathbf{c}_r\}_{r\in\mathcal{R}}$ of recorded clutter \textit{slow-time} vectors for each range bin in $\mathcal{R}$ is created by cropping $K$ consecutive pulses from all range bins, at a random offset within a file.

\subsection{Pre-Processing}\label{subsec:pre_process_flow}
\textcolor{black}{In this work, the real-valued NNs were considered.}
The pre-processing is used to convert the complex-valued input signal $\mathbf{X}$ from \eqref{eq:meas_model_SCW} to a \textcolor{black}{real-valued} representation that is appropriate for a NN processing and target detection in range or Doppler and is composed of 3 steps:

\makeatletter
\renewcommand{\ALG@name}{Pre-Processing flow}
\makeatother
\renewcommand{\thealgorithm}{}
\begin{algorithm}[H]
\caption{}
  \begin{algorithmic}[1]
  \Statex \textbullet~\textbf{Input:} \textcolor{black}{$\mathbf{X}\in\mathbb{C}^{N\times K}$}, detection parameter $p\in\{\text{``range''},\text{``Doppler''}\}$
  \begin{enumerate}
    \item Reshape input: 
    \begin{align}
    &\mathbf{X}_0 = 
    \begin{cases}
        \mathbf{X}^T,&\;p=\text{``range''}\nonumber\\
        \mathbf{X},&\;p=\text{``Doppler''}\nonumber
    \end{cases}
    \end{align}
    \item Center features over rows:
    \begin{align}
        [\mathbf{X}_1]_{[\cdot,:]} &= [\mathbf{X}_0]_{[\cdot,:]} - \overline{\mathbf{x}}^T.\nonumber
    \end{align}
    \item Concatenate real and imaginary terms:
    \begin{align}
        \mathbf{Z}_0 = \left[\Re{\mathbf{X}_1},\;\Im{\mathbf{X}_1}\right]\;.\nonumber
    \end{align}
    \end{enumerate}
    \Statex \textbullet~\textbf{Output:} \textcolor{black}{$\mathbf{Z}_0=\mathcal{P}\left(\mathbf{X}\right)\in\{\mathbb{R}^{N\times 2K},\mathbb{R}^{K\times 2N}\}$}
  
  \end{algorithmic}
\end{algorithm}
The row vector, $[\mathbf{X}]_{[\cdot, :]}$, represents a row in $\mathbf{X}$ and $\quad\overline{\mathbf{x}}^T=\frac{1}{M_0}\sum_{m=1}^{M_0}{[\mathbf{X}_0]_{[m,:]}}$, where $M_0$ is the number of rows in $\mathbf{X}_0$.
Note that the target echo signal in \eqref{eq:target_echo_signal} is an outer product of two complex sinusoidal vectors, such that the range and Doppler information is encoded in the frequency content along the corresponding axis. 
\textit{Step 1} transposes the input frame $\mathbf{X}$ such that the column axis represents the parameter of interest (i.e. the ``feature axis'') and the row axis contains different realizations of that feature. 
\textit{Step 2} centers each feature and \textit{Step 3} concatenates the real and imaginary parts of the data \textcolor{black}{ to  real-valued representation, $\mathbf{Z}_0$, of dimensions $N\times 2K$, for $p=\text{``Doppler''}$, and $K\times 2N$ for $p=\text{``range''}$}.

\begin{figure*}[!t]
\centering
\includegraphics[width=0.8\textwidth]{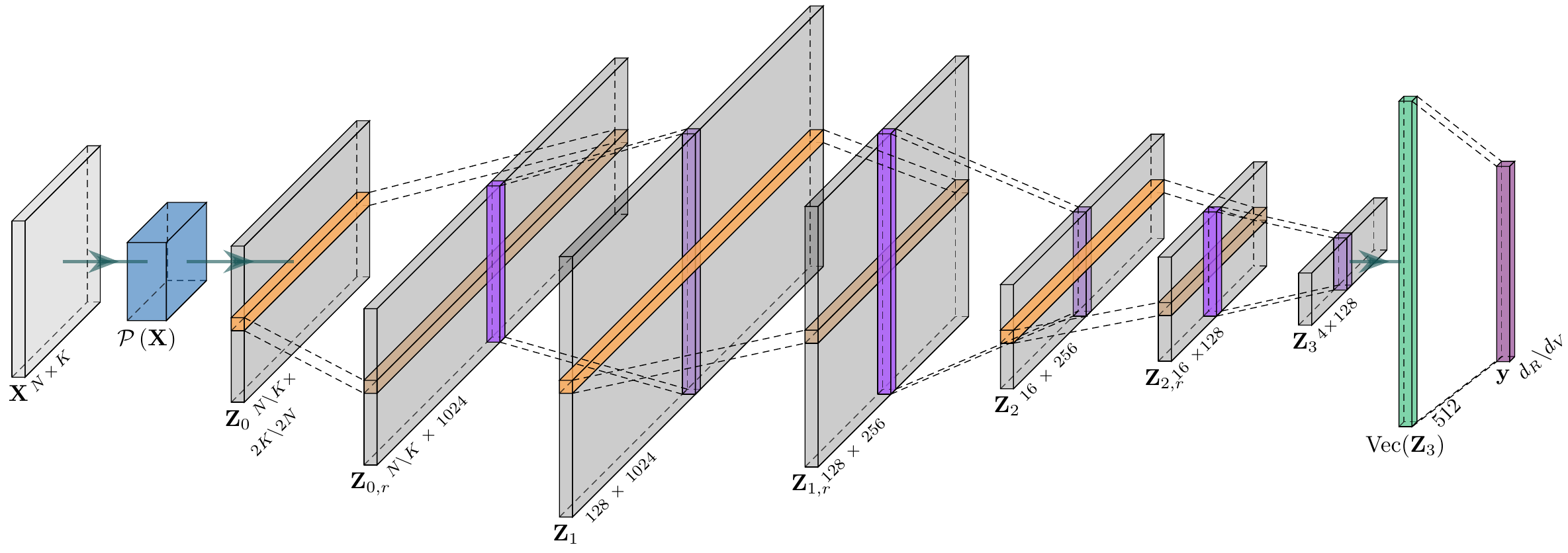}
\caption{Neural network architecture described in Subsection~\ref{subsec:nn_arch_and_training}. 
$\mathbf{X}$ is the \textcolor{black}{complex-valued}  \textit{fast-time} $\times$ \textit{slow-time} input frame, $\mathcal{P}\left(\mathbf{X}\right)$ is the pre-processing flow described in Subsection~\ref{subsec:pre_process_flow} and $\mathbf{Z}_0\triangleq\mathcal{P}\left(\mathbf{X}\right)$ \textcolor{black}{is the real-valued matrix output of the pre-processing flow}. The row mapping (orange) and column mapping (magenta) represent steps 1 and 2 from the DAFC in Subsection \ref{subsubsec:dafc_block} with \textit{tanh} activation function. The dashed lines represent the mapping operation via a FC transform. 
The final layer outputs the vector $\mathbf{y}$ which consists of sigmoid activation at each output neuron.}
\label{fig:sep_fc_nn}
\end{figure*}

\subsection{Dimensional-Alternating Fully Connected}\label{subsubsec:dafc_block}
This Subsection presents a novel neural processing block for radar target detection in range or Doppler domains.

\makeatletter
\renewcommand{\ALG@name}{Dimensional-Alternating Fully Connected}
\makeatother
\begin{algorithm}[H]
\caption{}
  \begin{algorithmic}[1]
  \Statex \textbullet~\textbf{Input:} $\mathbf{Z}_{in}\in\mathbb{R}^{H\times W}$
   \begin{enumerate}
   \item FC transform of each row in $\mathbf{Z}_{in}$:
    \begin{align}
        \mathbf{Z}_r=\mathcal{F}_{\mathbf{h}_r,\mathbf{W}_r,\mathbf{b}_r}\left(\mathbf{Z}_{in}\right)\triangleq h_r\left(\mathbf{Z}_{in}\mathbf{W}_r + \mathbf{1}_{H}\mathbf{b}_r^T \right)\nonumber
    \end{align}
    \item FC transform of each column in $\mathbf{Z}_{r}$:
    \begin{align}
        &\mathbf{Z}_{out}=\mathcal{G}_{\mathbf{h}_c,\mathbf{W}_c,\mathbf{b}_c}\left(\mathbf{Z}_r\right)\triangleq h_c\left(\mathbf{W}_c^T\mathbf{Z}_r + \mathbf{b}_c\mathbf{1}_{W'}^T\right)\nonumber
    \end{align}
    \end{enumerate}
    \Statex \textbullet~\textbf{Output:} $\mathbf{Z}_{out} = \mathcal{S}\left(\mathbf{Z}_{in}\right)\in\mathbb{R}^{H'\times W'}$
  \end{algorithmic}
\end{algorithm}
Let $\mathcal{F}_{\mathbf{h}_r,\mathbf{W}_r,\mathbf{b}_r}$ denote a fully connected (FC) transform applied to each row in $\mathbf{Z}_{in}$ and $\mathcal{G}_{\mathbf{h}_c,\mathbf{W}_c,\mathbf{b}_c}$ denote a FC transform applied to each column in $\mathbf{Z}_r$.
$\mathbf{W}_r\in\mathbb{R}^{W\times W'},\mathbf{b}_r\in\mathbb{R}^{W'},\mathbf{W}_c\in\mathbb{R}^{H\times H'},\mathbf{b}_c\in\mathbb{R}^{H'}$ represent the DAFC block's learnable parameters and $h_r\left(\cdot\right),h_c\left(\cdot\right)$ are nonlinear activation functions, which are applied element-wise. This block is repeatedly used in a pipeline structure, such that the input to the first block is the output of the pre-processing flow $\mathbf{Z}_0=\mathcal{P}\left(\mathbf{X}\right)$.

The DAFC block is designed according to the three following ideas: 1) \textbf{structured transformation}, 2) \textbf{sparsity}, and 3) \textbf{non-linearity}.

\textbf{1) Structured transformation} 

\textcolor{black}{The proposed radar signal processing approach is specifically tailored to the conventional radar with the  \textit{fast-time} $\times$ \textit{slow-time} data structure}. 
From~\eqref{eq:meas_model_SCW},~\eqref{eq:target_echo_signal}, and the pre-processing procedure described in Subsection~\ref{subsec:pre_process_flow}, each row in $\mathbf{Z}_0$ encodes the information about the parameter of interest (i.e. range or Doppler) in its frequency content. 
The variation between the rows in $\mathbf{Z}_0$ is a result of the interaction between multiple target echo signals and the statistical variation of the clutter and noise, and the random phase for each single target echo signal.

With that view of information structure in mind, $\mathcal{F}_{\mathbf{h}_r,\mathbf{W}_r,\mathbf{b}_r}$ is a ``feature-extracting'' transform that aims to extract the parameter of interest related features from each row, that will contribute to the detection capability of the NN.
$\mathcal{G}_{\mathbf{h}_c,\mathbf{W}_c,\mathbf{b}_c}$, is a ``re-calibration" transform that inserts nonlinear interaction between transformed rows, in analogy to the excitation stage in \cite{hu2018squeeze}. This interaction provides an additional degree of nonlinear processing that can learn to mitigate clutter-induced interference and enhance the contribution of the transformed rows.
The DAFC operation also resembles the two-parts separable convolution~\cite{howard2017mobilenets} where an image's channel dimension and spatial dimension are processed consecutively.

\textbf{2) Sparsity}

The data processed by the proposed approach is matrix-shaped.
The straight-forward approach to process this type of data is to vectorize each matrix and invoke a vanilla FC layer.
In contrast, the proposed approach performs a matrix-to-matrix transform by consecutively processing the data along the two axes (rows and columns) while using a sparse set of parameters. 
For example, consider a transform from $\mathbf{Z}_{in}\in\mathbb{R}^{H_1\times W_1}$ into $\mathbf{Z}_{out}\in\mathbb{R}^{H_2\times W_2}$ and assume for simplicity that $H_1=W_1=M_1$ and $H_2=W_2=M_2$.
The number of parameters in such a transform is:
\begin{align}
    &\text{FC: }H_2W_2(H_1W_1+1)= O(M_1^2M_2^2)\;,\\
    &\text{DAFC: }W_2(W_1+1) + H_2(H_1+1)= O(M_1M_2)\;.\nonumber
\end{align}
Notice that the DAFC's parameter complexity grows linearly with the data dimensions, compared with the quadratic parameter complexity growth of the FC layer.
Since NNs are typically optimized using gradient-based learning~\cite{Goodfellow-et-al-2016}, this lower parameter complexity significantly reduces the dimensionality of the NN's learnable parameter space and therefore can aid in the convergence of the gradient-based learning procedure.

In addition, the parameter complexity coincides with the computational complexity.
Therefore, the design of the DAFC block contributes to the applicability of the NN.
The lower complexity also enables higher dimensional representations, which can contribute to the capacity~\cite{Goodfellow-et-al-2016} of the NN, hence enabling the NN to represent more complex and abstract representations that will enhance the detection capabilities.

\textbf{3) Non-linearity}

Another interesting attribute of the DAFC block, is the additional degree of non-linearity. The straight-forward FC-based approach will process the input matrix via an affine transformation that is followed by a nonlinear activation function. In contrast, the DAFC inserts an additional nonlinear activation after applying an affine map on each row. Although the affine maps in the DAFC block are of lower dimension, the additional degree of non-linearity can contribute to the capacity~\cite{Goodfellow-et-al-2016} of the NN, thus enabling more abstract and complex representations to be learned.

\begin{figure*}[!t]
\centering
\includegraphics[width=0.7\textwidth]{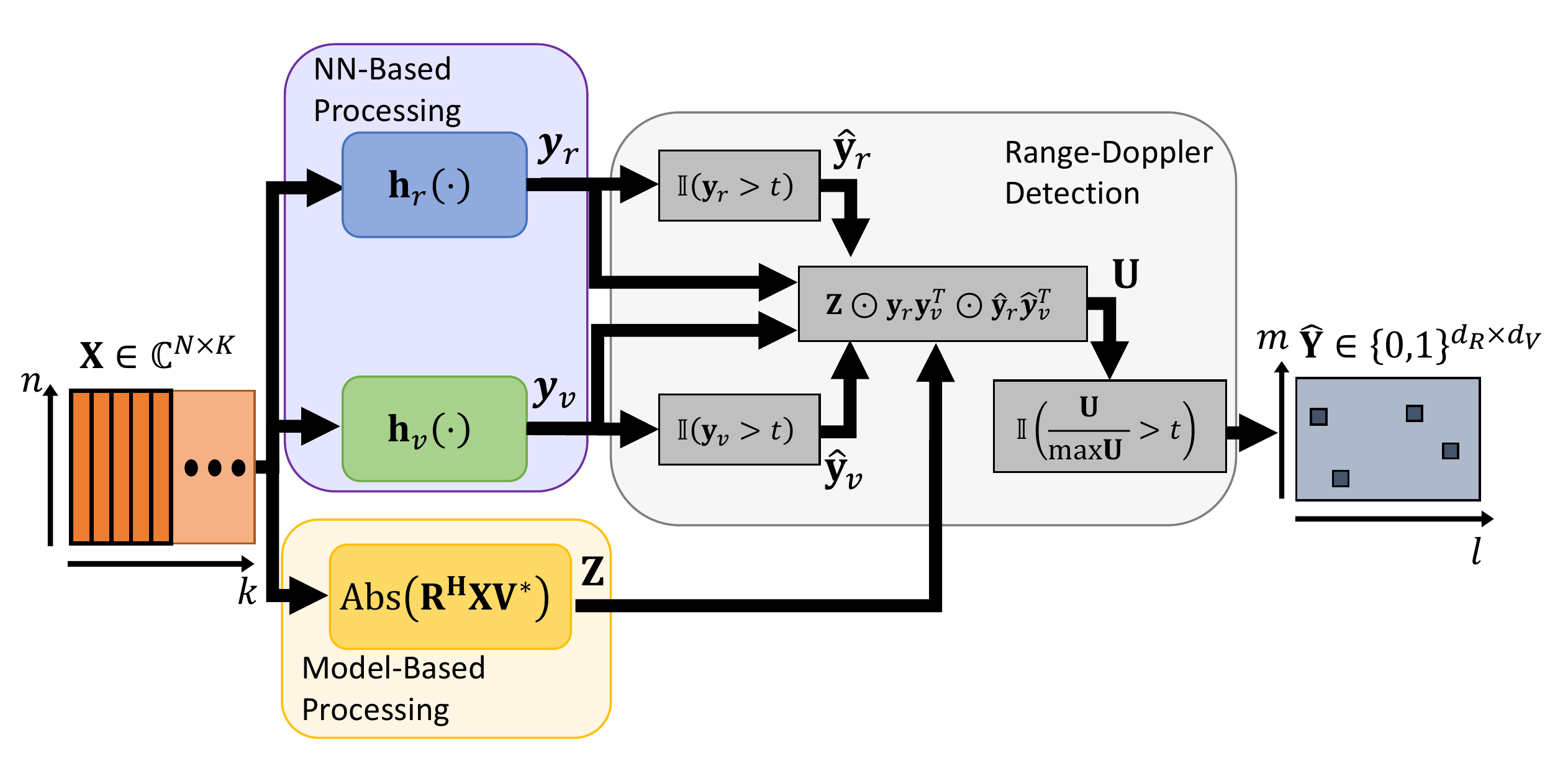}
\caption{Range-Doppler detection scheme.
$\mathbf{X}$ is the complex input signal. 
The NN-based processing (light purple) contains separate instances of NN architectures described at Subsection \ref{subsec:nn_arch_and_training} for range and Doppler.
The model-based processing projects the inputs signal onto the steering matrices as described in Subsection \ref{sec:rd_detection}.
The range-Doppler detection (light gray) contains the integration scheme between the NN and model-based outputs as detailed in Subsection \ref{sec:rd_detection} and provides the range-Doppler decision matrix $\hat{\mathbf{Y}}$.}
\label{fig:rd_detection}
\end{figure*}

\subsection{The NN Architecture and Training}\label{subsec:nn_arch_and_training}
The proposed NN architecture is selected to provide a mapping from the complex \textit{fast-time} $\times$ \textit{slow-time} input to a probability mass function (pmf) of a target presence in each range or Doppler bin.
To that aim, the pre-processing flow described in Subsection~\ref{subsec:pre_process_flow} and \textcolor{black}{the} DAFC neural processing described in Subsection~\ref{subsubsec:dafc_block} are utilized to compose a NN architecture that delivers the desired mapping.
The selected NN architecture is formulated as:
\begin{align}
&\mathbf{y}=\mathbf{h}\left(\mathbf{X}\right)\triangleq\mathcal{F}\left(\text{Vec}\left(\mathcal{S}_3\left(\mathcal{S}_2\left(\mathcal{S}_1\left(\mathcal{P}\left(\mathbf{X}\right)\right)\right)\right)\right)\right)\;,
\end{align}
where $\mathcal{P}$ is the pre-processing procedure described in Subsection \ref{subsec:pre_process_flow}, $\mathcal{S}_b$ denotes the enumerated $b$-th DAFC block and $\mathcal{F}$ is a FC transform (i.e. layer) with sigmoid activation at each output neuron.
Fig. \ref{fig:sep_fc_nn} shows the proposed NN architecture with parameters specified in Table \ref{tab:arch}.

\begin{table}[ht]
\begin{center}
 \begin{tabular}{m{2.5cm} m{2.5cm} m{2cm}}
 \hline
 Block & Output Dimension & Activation\\ [1.0ex] 
 \hline\hline
 $\mathbf{Z}_0=\mathcal{P}\left(\mathbf{X}\right)$ & $K\times 2N$ or $N\times 2K$ & - \\ [1.0ex]
 \hline
 $\mathbf{Z}_1=\mathcal{S}_1\left(\mathbf{Z}_0\right)$ & $128\times 1024$ & tanh \\ [1.0ex]
 \hline
$\mathbf{Z}_2=\mathcal{S}_2\left(\mathbf{Z}_1\right)$ & $16\times 256$ & tanh \\ [1.0ex]
 \hline
$\mathbf{Z}_3=\mathcal{S}_3\left(\mathbf{Z}_1\right)$ & $4\times 128$ & tanh \\ [1.0ex]
 \hline
 Vec$\left(\mathbf{Z}_3\right)$ & 512 & - \\ [1.0ex]
 \hline
 $\mathbf{y}=\mathcal{F}\left(\text{Vec}\left(\mathbf{Z}_3\right)\right)$ & $d_R$ or $d_V$ & sigmoid \\ [1.0ex]
 \hline
\end{tabular}
\end{center}
\caption{\label{tab:arch} NN architecture. 
Each row contains the specifications of each block in the NN.
\textcolor{black}{The number of parameters in the NN for $N=K=64$ and $d=d_R=d_V=64$ is 470,676.}}
\end{table}

The motivation for the dimensionality expansion in the early stages is similar to the high-dimensional feature space rationale used in kernel-SVM methods \cite{shalev2014understanding}. 
Namely, the NN learns a high-dimensional mapping that transforms the input data to a high-dimensional space, in which the expressive characteristics are enhanced and the detection ability is increased.
The dimensionality reduction in later stages is aimed to reduce the dimension of the latent representation to a dimension that is closer to the label space dimension.
This trend of dimensionality expansion by early layers and reduction in later layers is also presented in \cite{ansuini2019intrinsic}, where the authors show that the intrinsic dimensionality of CNNs' layers follows the same trend.

The last activation is chosen to be sigmoid since we choose the NN to produce a pmf of target presence in each range or Doppler bin. Thus, the multi-dimensional pmf, $\mathbf{y}\in[0,1]^d$, where $d\in\{d_R,d_V\}$, encodes the target presence probability in each bin. 
The proposed NN architecture enables detection of \textbf{multiple targets} at a single inference cycle, without the need for ROI pooling of range or Doppler subspaces, similarly to the object detection approach in~\cite{redmon2018yolov3}. Similarly to \cite{brodeski2019deep}, this architecture considers a sparse label. Therefore, the class-balanced cross-entropy~\cite{cui2019class} is used as the following loss function:
\begin{align}\label{eq:loss_function}
    \mathcal{L}\left(\mathbf{y},\mathbf{y}_{true}\right) = -\frac{1}{d}\sum_{j=0}^{d-1}&\frac{1-\beta}{1-\beta^{n_0}}\left(1 - [\mathbf{y}_{true}]_j\right)\log{\left(1 - [\mathbf{y}]_j\right)}\\
    &+ \frac{1-\beta}{1-\beta^{n_1}}\left([\mathbf{y}_{true}]_j\right)\log{\left([\mathbf{y}]_j\right)}\;,\nonumber
\end{align}
where $\beta\rightarrow1$, $\mathbf{y}\in[0,1]^d$ and $\mathbf{y}_{true}\in\{0,1\}^d$ are the NN output and true label vector, respectively, $d$ is the dimension of label vector ($d_R$ or $d_V$). Terms $n_1$, $n_0$ are proportional to the number of the target bins and the target-free bins, respectively, and represent density of the target labels within the training data set. 

\subsection{NN-Based Range-Doppler Detector}\label{sec:rd_detection}
Let the range and Doppler steering matrices be: $\mathbf{R} = 
    \begin{bmatrix} 
        \mathbf{r}(0) & \mathbf{r}(\Delta r) & \dots & \mathbf{r}(\Delta r (d_R-1))
    \end{bmatrix}$ 
    and $ \mathbf{V} = 
    \begin{bmatrix} 
        \mathbf{v}(0) & \mathbf{v}(\Delta v) & \dots & \mathbf{v}(\Delta v (d_V-1))
    \end{bmatrix}$,
where $\mathbf{r}(r)$ and $\mathbf{v}(v)$ are defined in \eqref{eq:target_echo_signal_parameters}.
Let $\mathbf{h}_r(\cdot)$ and $\mathbf{h}_v(\cdot)$ denote the range and Doppler NN models, which are in fact separate instances of the NN described in Subsection~\ref{subsec:nn_arch_and_training}.
Fig. \ref{fig:rd_detection} shows the range-Doppler detection scheme detailed in \textbf{Algorithm 1}.

\makeatletter
\renewcommand{\ALG@name}{Algorithm 1: Range-Doppler Detection}
\makeatother
\begin{algorithm}[H]
\caption{}
  \begin{algorithmic}[1]
  \Statex \textbullet~\textbf{Input:} $\mathbf{X}\in\mathbb{R}^{N\times K}$, detection threshold $t$
  \begin{enumerate}
    \item  NN feed-forward, range-Doppler projection
    \begin{align}
        \mathbf{y}_r=\mathbf{h}_r\left(\mathcal{\mathbf{X}}\right), \;\mathbf{y}_v=\mathbf{h}_v\left(\mathcal{\mathbf{X}}\right),\;\mathbf{Z}=\text{Abs}\left(\mathbf{R}^H\mathbf{X}\mathbf{V}^*\right)\nonumber
    \end{align}
    \item Apply threshold to NN-yielded pmf's and use them to filter the projected matrix:
    \begin{align}
        \hat{\mathbf{y}}_r &= \mathbb{I}\left(\mathbf{y}_r > t\right),\;\hat{\mathbf{y}}_v = \mathbb{I}\left(\mathbf{y}_v > t\right)\;,\nonumber\\
        \mathbf{U}&= \mathbf{Z}\odot\mathbf{y}_r\mathbf{y}_v^T\odot\hat{\mathbf{y}}_r\hat{\mathbf{y}}_v^T\nonumber
    \end{align}
    \item Apply threshold to obtain range-Doppler detection matrix:
    \begin{align}
        \hat{\mathbf{Y}} = \mathbb{I}\left(\frac{1}{\text{max}\mathbf{U}}\mathbf{U} > t\right)\nonumber
    \end{align}
    \end{enumerate}
    \begin{itemize}
    \item \textbf{Output:} $\hat{\mathbf{Y}}\in\mathbb{R}^{d_R \times d_V}$
    \end{itemize}
  \end{algorithmic}
\end{algorithm}
\noindent In \textit{Step 1}, $\mathbf{y}_r$ and $\mathbf{y}_v$ are NN-yielded multi-dimensional pmf's of target presence in each range and Doppler bin, and $\mathbf{Z}$ is the projection of the input frame on the steering vectors representing the range and Doppler bins.
\textit{Step 2} uses $\mathbf{y}_r$ and $\mathbf{y}_v$ to detect targets, mitigate clutter and factorize $\mathbf{Z}$ at each range-Doppler bin to obtain $\mathbf{U}$.
In \textit{Step 3}, $\mathbf{U}$ is normalized to values in the interval $[0,1]$, such that the detection threshold $t$ can be used. 
Note that $\mathbf{U}$ is the combination of the model-based transform (i.e. projection to range-Doppler steering matrices, which is equivalent to conventional range-Doppler transform), NN-based transform (to obtain a $[0,1]$ score for each range-Doppler bin in $d_R\times d_V$), and range-Doppler bins, which exceed the threshold, $t$, at each NN output (i.e. targets detected by the NNs). 

\textcolor{black}{The detection threshold $t$ determines the trade-off between the probability of detection, $P_D$, and the probability of false-alarm  $P_{FA}$. In this work, the threshold is set empirically for the desired $P_{FA}$.}
\textcolor{black}{Although the CFAR property of the proposed approach is not theoretically guaranteed,  the experiment in Subsection~\ref{subsubsec:pfa_miss} shows that the proposed approach is robust to variation in $\nu$ in terms of false-alarm probability, compared to the CA-CFAR and \textcolor{black}{TM-CFAR} detectors.}

The range-Doppler detector in \textbf{Algorithm 1} contains a combination of NN-based processing and model-based processing.
The goal of the NNs is to output a pmf for the presence of target in each range or Doppler bin ($\mathbf{y}_r$ and $\mathbf{y}_v$), that attains low probability for clutter-containing and/or target-free bins, whereas high probability for target-containing bins. 
Therefore, the proposed NN has the ability to mitigate the clutter in the detector output, since low-probability bins are filtered by the detection threshold $t$, whereas the model-based projection is not able to suppress the correlated heavy-tailed clutter's energy, as explained in Subsection~\ref{subsec:rd_detection_formulation}.
On the other hand, solely relying on the NNs as range-Doppler detectors suffers from an inherent ambiguity in the combined range-Doppler space, since the operation $\mathbf{y}_r\mathbf{y}^T_v$ results in a Cartesian product of all possible range-Doppler combinations. To that aim, the model-based projection delivers a signal that contains information regarding energy presence in each range-Doppler bin, and thus compensating for the NN-based ambiguity.
The experiment described in Subsection \ref{subsec:eval_combination_nn_sp}, shows the significant gain obtained by this combination.


\section{PERFORMANCE EVALUATION}\label{sec:eval}

For all experiments, we have used synthetic simulated targets according to the target signal model described in Subsection~\ref{subsec:meas_model}.
A training dataset of $10,000$ frames containing targets ($\mathcal{T}\ne\emptyset$) and $10,000$ frames without targets ($\mathcal{T}=\emptyset$), was used in each experiment to keep a balanced training dataset. 
For each target-containing frame (i.e. $\mathcal{T}\ne\emptyset$) the number of targets is $|\mathcal{T}|\sim\text{Unif}\left(\{1,\dots,8\}\right)$ with targets' parameters (range and Doppler) sampled uniformly over the continuous range-Doppler space:
\begin{align}\label{eq:target_rd_dist}
    r\sim\text{Unif}\left([r_{min},r_{max}]\right),\;v\sim\text{Unif}\left([v_{min},v_{max}]\right)\;.
\end{align}
For each target in $\mathcal{T}$, the SCNR is sampled from $\text{SCNR}_{[dB]}\sim\text{Unif}\left(-5,10\right)$.
In order to enrich the training dataset, each batch is simulated independently to increase the number of frames for each NN training.
Adam optimizer~\cite{kingma2014adam} with learning rate $10^{-3}$ and $\beta=0.99$ parameter were  used to train the network, together with $L_2$ regularization factor of $5\cdot10^{-4}$. 
Batch size of $256$, $300$ epochs, and a plateau learning rate scheduler with $0.905$ factor were selected.

\textcolor{black}{The test dataset in each evaluation scenario contains $4,000$ \textit{fast-time}$\times$\textit{slow-time} frames with $\mathcal{T}\ne\emptyset$ and $2,000$ frames with $\mathcal{T}=\emptyset$. 
For each frame with $\mathcal{T}\ne\emptyset$, the number of targets is, unless stated else, $4$, with equal $\text{SCNR}$ for each target.
Range and Doppler values of targets are sampled in the same manner as the training dataset.}
The rest of the datasets' parameters are detailed in Table~\ref{tab:datasets_params}.

\begin{table}[ht]
\begin{center}
 \begin{tabular}{m{1.5cm} m{2.5cm} m{3.5cm}}
 \hline
 Notation & Description & Value\\ [1.0ex] 
 \hline\hline
 $B$ & Chirp bandwidth & $50\;[MHz]$\\ [1.0ex]
 \hline
 $T_0$ & PRI & $1\;[msec]$\\ [1.0ex]
 \hline
 $N$ & Samples per pulse (\textit{fast-time}) & $64$\\ [1.0ex]
 \hline
 $K$ & Pulses per frame (\textit{slow-time}) & $64$\\ [1.0ex]
 \hline
 $f_c$ & Carrier frequency& $9.39\;[GHz]$\\ [1.0ex]
 \hline
 $\sigma_f^2$ & Inverse of clutter corr. coefficient& $0.05^2$\\ [1.0ex]
 \hline
 $[r_{min},r_{max}]$ & Range interval & $[0,93]\;[m]$\\ [1.0ex]
 \hline
 $[v_{min},v_{max}]$ & Doppler interval& $[-7.5,7.5]\;[m/sec]$\\ [1.0ex]
 \hline
 $\mathcal{R}$ & Range bins& $\{0,3,\dots,93\}[m]$\\ [1.0ex]
 \hline
 $\Delta r$ & Range resolution& $3\;[m]$\\ [1.0ex]
 \hline
 $\mathcal{V}$ & Doppler bins& $\{-7.73,-7.48,\dots,7.73\}$ $[m/sec]$\\ [1.0ex]
 \hline
 $\Delta v$ & Doppler resolution& $0.249\;[m/sec]$\\ [1.0ex]
 \hline
 $d_{R}=|\mathcal{R}|$ & Number of range bins& $32$\\ [1.0ex]
 \hline
 $d_{V}=|\mathcal{V}|$ & Number of Doppler bins& $63$\\ [1.0ex]
 \hline
 CNR & Clutter-to-noise ratio & $15$ [dB]\\ [1.0ex]
 \hline

\end{tabular}
\end{center}
\caption{\label{tab:datasets_params} Parameters of generated datasets.}
\end{table}

In the following subsections, the performance of the proposed approach is evaluated in a clutter-free scenario and in various scenarios with both simulated and recorded radar clutter. 
The evaluation is via the probability of detection ($P_D$) for a fixed probability of false-alarm ($P_{FA}$).
\textcolor{black}{The performance on $30$ independently-generated test datasets are averaged in each experiment in order to display the results.}
For each generated dataset, the detection threshold is set to determine a predefined $P_{FA}$.
Note that in each experiment, \textbf{the same} range NN instance and \textbf{the same} Doppler NN instance are used for various SCNRs and various clutter conditions, as further detailed in the experiments below.

\textcolor{black}{
In the following experiments, the performance of the proposed approach was evaluated and compared to a) ANMF, on {\it slow-time} signals at each range bin and b) CA-CFAR and \textcolor{black}{TM-CFAR}, on the 2D FFT-based range-Doppler energy map.
}
\textcolor{black}{
The NMF~\cite{conte1995asymptotically} is the GLRT approximation for scenarios with a single target within a CUT. However, NMF requires the true clutter covariance matrix and therefore, is impractical.
The ANMF, is the adaptive NMF that utilizes the estimated clutter covariance matrix~\cite{coluccia2021knn}.
The $\Sigma$-ANMF~\cite{conte1998adaptive,coluccia2021knn} estimates the clutter covariance matrix using the $\Sigma$ estimator and therefore, is suited for scenarios with non-Gaussian clutter~\cite{conte1998adaptive}.}

\textcolor{black}{
CA-CFAR is an optimal detector for targets within a homogeneous environment in the range-Doppler domain, and the \textcolor{black}{TM-CFAR} is a robust CFAR method that is designed to operate in a heterogeneous environment in the range-Doppler domain~\cite{richards2010principles}. 
The selected window size is $9\times 15$ with $3\times3$ guard cells for the CA-CFAR \textcolor{black}{and TM-CFAR detectors}. 
Similarly to the proposed method, the detection threshold for the $\Sigma$-ANMF, CA-CFAR, and \textcolor{black}{TM-CFAR} is determined empirically according to the desired $P_{FA}$.
}


\subsection{Performance Evaluation Metrics}\label{subsec:eval_metric}
The metrics for the following performance evaluation, $P_D$ and $P_{FA}$, are defined in this Subsection.
Consider a dataset generated according to the guidelines in Subsection~\ref{sec:data_gen}, and let $(i,j)$ denote the index of the $j$-th target in the $i$-th example in the dataset.
For each target $\left(r_{i,j},v_{i,j}\right)$ denote the indices of the ``closest" range and Doppler bins by $[m_{i,j},l_{i,j}]$. Define the ``neighboring" box $B\left(\left[m_{i,j},l_{i,j}\right]\right)$ as:
\begin{align}
    B\left(\left[m_{i,j},l_{i,j}\right]\right) = [m_{i,j},l_{i,j}] \bigcup \left\{\left([m_{i,j}\pm 1,l_{i,j} \pm 1]\right)\right\}\;.
\end{align}
The event of successfully detecting the target at $\left[m_{i,j},l_{i,j}\right]$ is defined as:
\begin{align}\label{eq:sucsessful_detection}
    D\left(\left[m_{i,j},l_{i,j}\right]\right) = \mathbb{I}\left( \exists[m,l]\in B\left(\left[m_{i,j},l_{i,j}\right]\right) : [\hat{\mathbf{Y}}]_{m,l}=1 \right)\;.
\end{align}
In words, we define successful target detection if the true target is located at most one bin away from a predicted target position.
The probability of detection, $P_D$, is evaluated by the ratio between the number of successfully detected targets, according to (\ref{eq:sucsessful_detection}), to the total number of targets in the dataset:
\begin{align}
    P_D = \frac{1}{\left|\bigcup_{i,j}{\left(r_{i,j},v_{i,j}\right)}\right|}\sum_{i,j}{D\left(\left[m_{i,j},l_{i,j}\right]\right)}\;.
\end{align}
\textcolor{black}{Correspondingly, the $P_{FA}$ is evaluated as the number of false-detected range-Doppler bins:
\begin{align}
    P_{FA} &= \frac{1}{\left|B_0\right|}\sum_{[m,l]\in B_0}{D\left(\left[m,l\right]\right)}\;,\\
    B_0 &= \bigcup_{i}{\left(\{1,\dots,d_R\}\times \{1,\dots,d_V\} \setminus \bigcup_{j}{B\left(\left[r_{i,j},v_{i,j}\right]\right)}\right)}\;,\nonumber
\end{align}
}
where $B_0$ is the ``no-target" space, defined as the set of indices of all range-Doppler bins in the dataset, excluding the true targets and their corresponding ``neighboring'' boxes.

\subsection{Clutter-Free Scenario}
In this scenario, the performance of the proposed approach is evaluated in a clutter-free environment.
In the clutter-free, single-target scenario, CA-CFAR is an approximation of the GLRT~\cite{richards2010principles}. \textcolor{black}{The \textcolor{black}{TM-CFAR} is designed for multi-target scenario and heterogeneous environments~\cite{richards2010principles}}.
The NN models in the proposed method are trained using data with half of the training frames containing AWGN ($\mathbf{C}=\mathbf{0}$) and half containing clutter, simulated with shape parameter $\nu$, sampled from $\nu\sim\text{Unif}\left(0.1,1.5\right)$.

\begin{figure}[ht]
\centering
\includegraphics[width=0.42\textwidth]{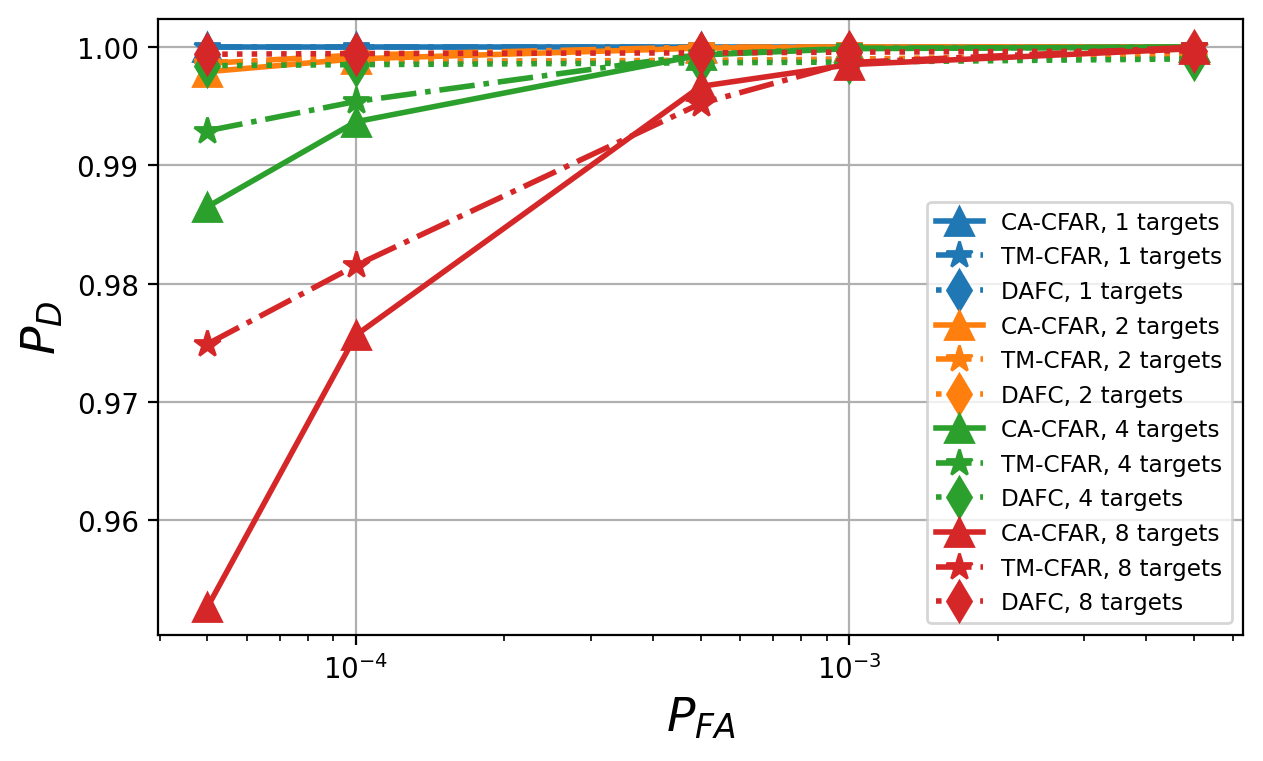}
\caption{ROC of evaluated detectors in clutter-free scenarios with $\{1,2,4,8\}$ targets and $SCNR = 0\;dB$.}
\label{fig:WGN}
\end{figure}

Fig. \ref{fig:WGN} compares the receiver operating characteristics (ROC) of the proposed approach to \textcolor{black}{the conventional CA-CFAR and \textcolor{black}{TM-CFAR} detectors} in scenarios with $\{1,2,4,8\}$ targets. Notice that in this scenario, the optimal single-target CA-CFAR detector attains $P_D=1$ for all $P_{FA}$ values. However, with increasing the number of targets, its $P_D$ is degraded as a result of mutual target masking~\cite{richards2010principles}, which is more prominent at lower $P_{FA}$. 
\textcolor{black}{TM-CFAR slightly outperforms CA-CFAR in multiple targets scenarios, as it discards the neighbouring targets from the averaging window~\cite{richards2010principles}.}
The proposed approach is robust to variation in the number of targets compared to the CA-CFAR and \textcolor{black}{TM-CFAR} detectors. 
Furthermore, the NN models used in this experiment were trained using a dataset that contains also correlated heavy-tailed clutter. This evidence demonstrates the generalization capability of the proposed NN-based detection approach to variation in the interference behavior.

\begin{figure}[ht]
\centering
\includegraphics[width=0.42\textwidth]{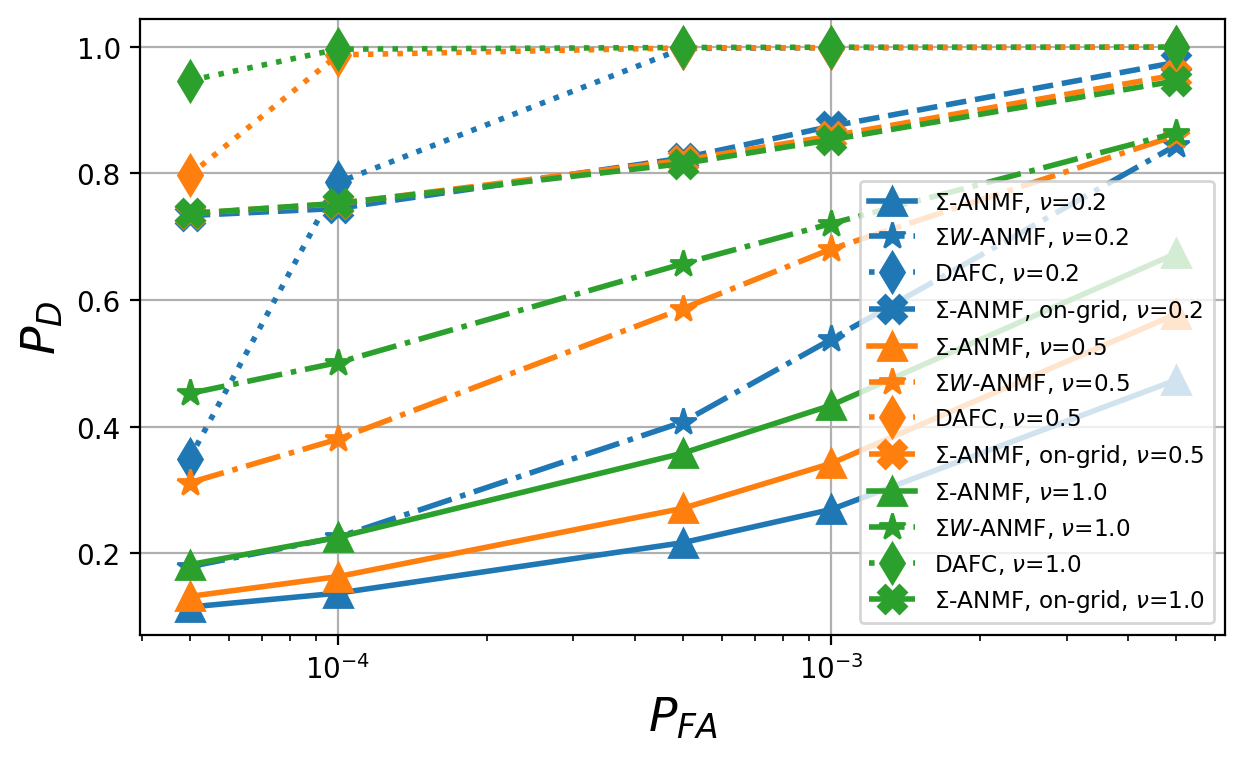}
\caption{ROC of multiple targets with simulated clutter, $\text{SCNR}=0\;dB$.}
\label{fig:anmf}
\end{figure}

\begin{figure*}[!t]
  \begin{subfigure}{0.42\textwidth}
    \caption{Clutter with targets, $P_{FA}=5 \cdot 10^{-4}$}
    \includegraphics[width=\linewidth]{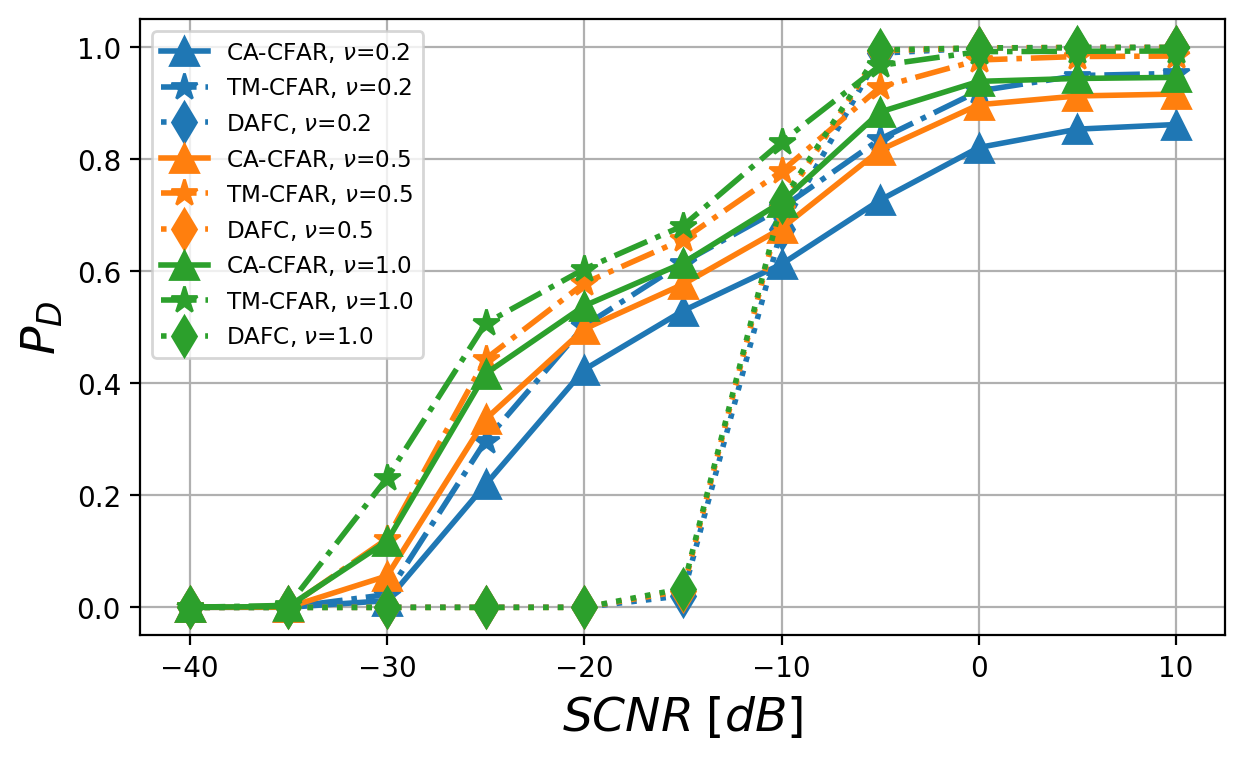}
    \label{subfig:sim_clutter_not_embedded_scnr}
  \end{subfigure}%
  \hspace*{\fill}   
  \begin{subfigure}{0.42\textwidth}
    \caption{Clutter with targets, $\text{SCNR}=0\;dB$}
    \includegraphics[width=\linewidth]{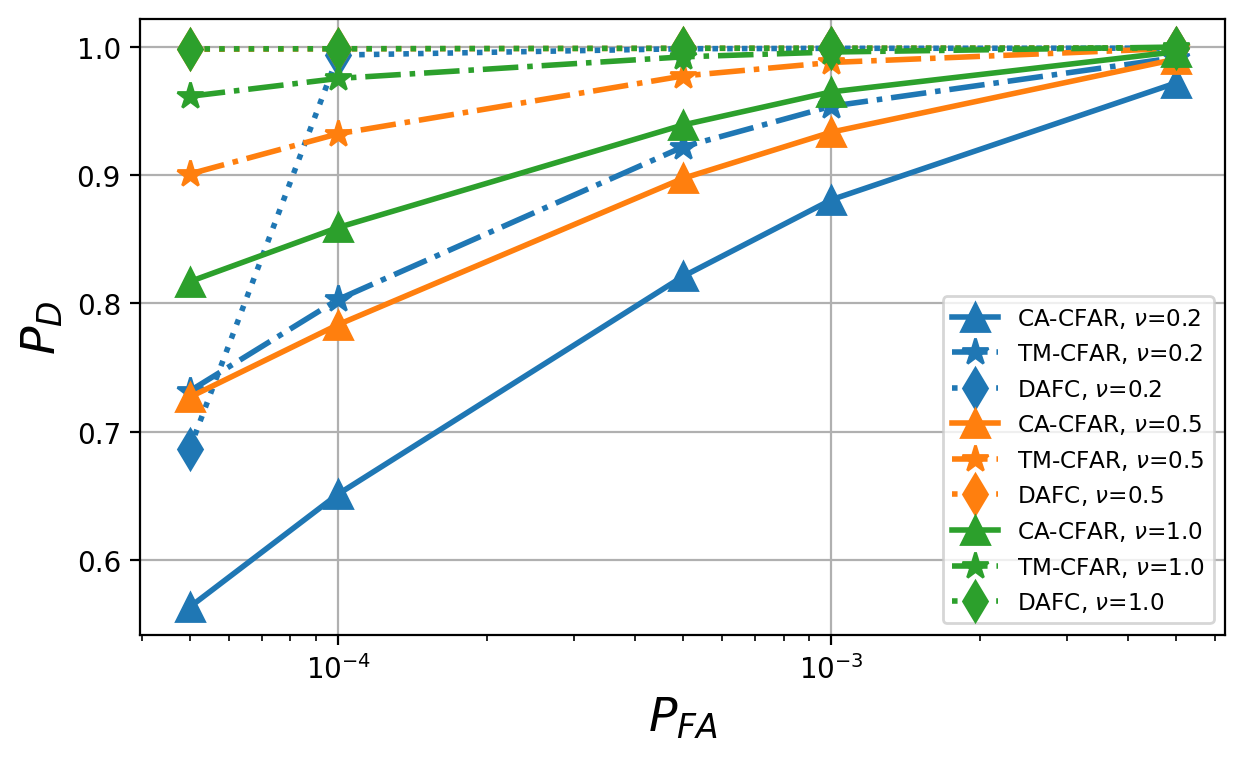}
    \label{subfig:sim_clutter_not_embedded_pfa}
  \end{subfigure}%
\caption{Detection results with simulated clutter with targets for various clutter ``spikiness" $\nu\in\{0.2,0.5,1.0\}$. Subplot (a) considers constant $P_{FA}=5\cdot10^{-4}$ and subplot (b) considers SCNR $=0\;dB$.} 
\label{fig:pd_sim_clutter}
\end{figure*}

\begin{figure*}[!t]
  \begin{subfigure}{0.42\textwidth}
    \caption{Clutter with embedded targets, $P_{FA}=5 \cdot 10^{-4}$}
    \includegraphics[width=\linewidth]{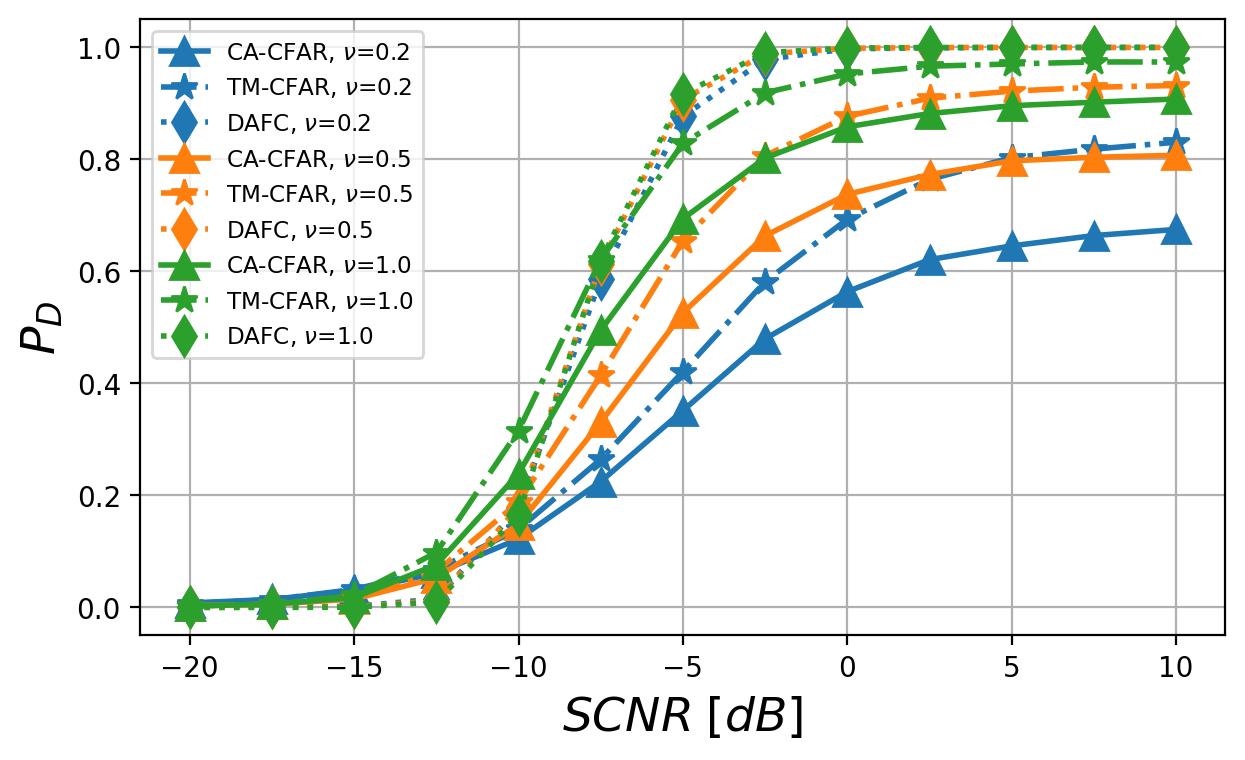}
    \label{subfig:sim_clutter_embedded_scnr}
  \end{subfigure}%
  \hspace*{\fill}   
  \begin{subfigure}{0.42\textwidth}
    \caption{Clutter with embedded targets, $\text{SCNR}=0\;dB$}
    \includegraphics[width=\linewidth]{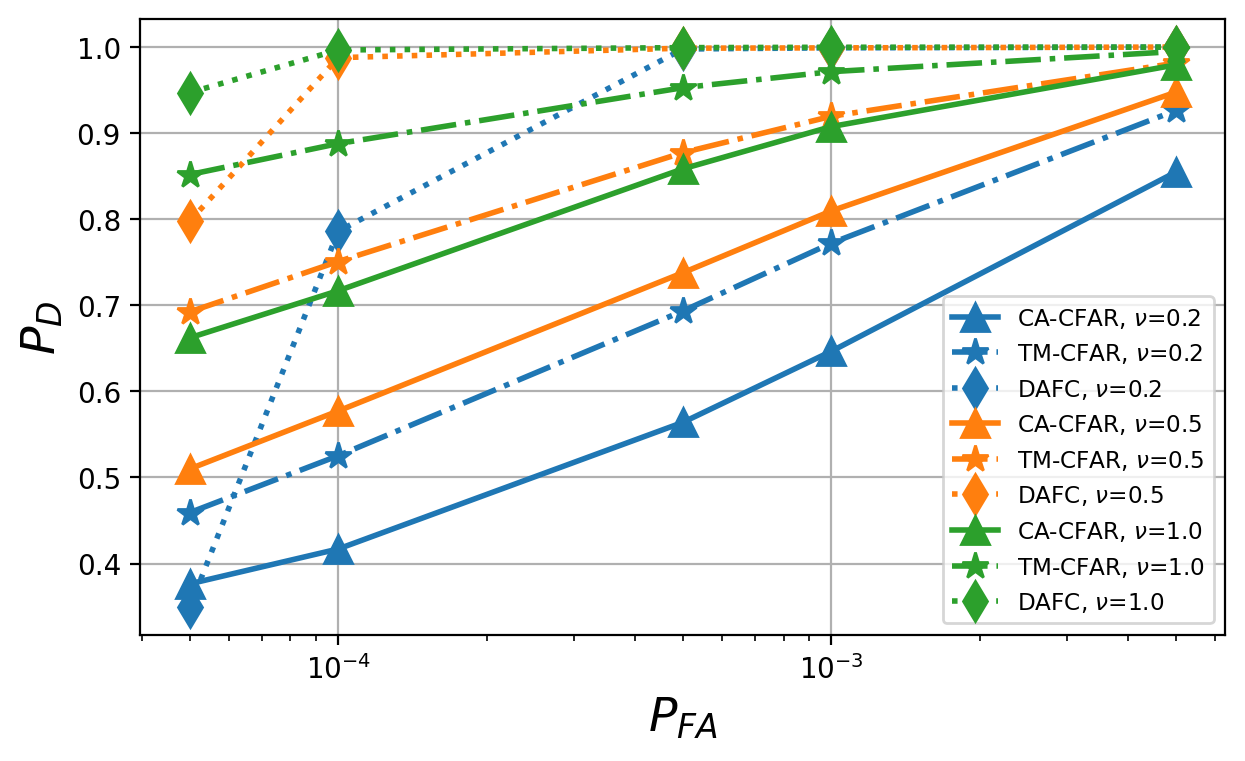}
    \label{subfig:sim_clutter_embedded_pfa}
  \end{subfigure}%
\caption{Detection results with simulated clutter with embedded targets for various clutter ``spikiness" $\nu\in\{0.2,0.5,1.0\}$. Subplot (a) considers constant $P_{FA}=5\cdot10^{-4}$ and subplot (b) considers SCNR $=0\;dB$.} 
\label{fig:pd_sim_clutter_embedded}
\end{figure*}

\subsection{Simulated Clutter}\label{subsec:eval_simulated_clutter}
In this experiment, the performance of the proposed approach is evaluated in scenarios with simulated correlated heavy-tailed clutter, as described in Subsection~\ref{sec:data_gen}. 
For the proposed approach, the NN models were trained using training sets with shape parameter $\nu$ sampled from $\nu\sim\text{Unif}\left(0.1,1.5\right)$. 
The evaluation is performed using three different test sets, generated with shape parameters, $\nu\in\{0.2,\;0.5,\;1.0\}$.

\subsubsection{Comparison to ANMF}
\textcolor{black}{
The ANMF detector is the GLRT approximation for a single target within SIRV clutter. For $\mathcal{H}_1$ hypothesis, it considers a single target within the range CUT and the availability of target-free secondary data.
Therefore, the ANMF is implemented via a sequence of binary hypothesis tests per range bin by scanning the Doppler values, $\mathcal{V}$. The range-Doppler bins with energy exceeding the detection threshold are declared as containing detected targets.
In this work, we consider a modification of the $\Sigma$-ANMF detector, denoted as $\Sigma W$-ANMF, which aims to suppress target-presence in neighbouring range cells.
The $\Sigma W$-ANMF involves Hanning window applied to each \textit{fast-time} signal, and taking $4$ guard cells around the range CUT, prior to the range-transform and $\Sigma$-ANMF.
Targets' Doppler values are drawn from a continuous interval, and thus, the binary hypothesis test formulation is mismatched in the desired \textit{slow-time} signal model. 
}

\textcolor{black}{Fig.~\ref{fig:anmf} shows the ROC of the proposed approach compared to the $\Sigma$-ANMF~\cite{conte1998adaptive} \textcolor{black}{for targets that are simulated with Doppler values in at most $\pm$ $1.5$ $m/sec$ offset from the clutter's Doppler, denoted as $f_d$ in \eqref{eq:cg_clutter}.}
It shows that in on-gird scenarios, where the target range and Doppler values are drawn from $\mathcal{R}\times\mathcal{V}$, the $\Sigma$-ANMF performance is significantly higher comparing to the scenarios where the target range and Doppler values do not exactly match the range-Doppler gird. In these scenarios, the high range side-lobes may compromise the target-free secondary data assumption.
\textcolor{black}{Fig.~\ref{fig:anmf} shows that the proposed approach significantly outperforms the $\Sigma$-ANMF detector, since the measurement model of the ANMF is mismatched in the considered here scenario.}
}

\subsubsection{Detection Performance}\label{subsubsec:eval_sim_clutter_performance}
Using the simulated correlated heavy-tailed clutter, the performance of the proposed detection approach is evaluated in a) clutter with targets and b) clutter with embedded targets scenarios.
In the first scenario, the Doppler of targets and clutter is independent, whereas in the second scenario the targets' Doppler are randomly set to be at most $\pm 1.5$ $m/sec$ offset from the clutter's Doppler, represented by $f_d$ in~\eqref{eq:cg_clutter}.
Figs. \ref{fig:pd_sim_clutter} and \ref{fig:pd_sim_clutter_embedded} show the $P_D$ of the proposed approach compared to the CA-CFAR \textcolor{black}{and \textcolor{black}{TM-CFAR}} detectors in scenarios with simulated heavy-tailed clutter with $\nu\in\{0.2,\;0.5,\;1\}$ as a function of SCNR and $P_{FA}$ for the embedded and not embedded scenarios.

The parameter $\nu$ introduced in the model in Subsection \ref{sec:data_gen}, controls the ``spikiness'' of the simulated clutter amplitude. 
 Figs. \ref{fig:pd_sim_clutter} and \ref{fig:pd_sim_clutter_embedded} show the robustness of the proposed approach to the ``spikiness'' of the simulated clutter.
\textcolor{black}{In both scenarios, TM-CFAR outperforms CA-CFAR due to its improved ability to address the inhomogeneous environments~\cite{richards2010principles}.}
\textcolor{black}{
In Fig.~\ref{fig:pd_sim_clutter}, the performance of the proposed approach significantly degrades for very low SCNR.
However, for higher SCNR, the proposed approach is robust for variations in clutter spikiness (varying $\nu$), comparing to the conventional CA-CFAR and TM-CFAR detectors, whose performance significantly degrades in the presence of spiky clutter.
In addition, the $P_D$ of the conventional methods does not always achieve $P_D=1$ in high SCNR scenarios, due to high side-lobes of the strong targets and possible target masking.}
In both Figs.~\ref{fig:pd_sim_clutter} and ~\ref{fig:pd_sim_clutter_embedded}, the proposed approach outperforms the CA-CFAR and \textcolor{black}{TM-CFAR} detectors and shows a minor degradation between the homogeneous and inhomogeneous scenarios.
These observations demonstrate the ability of the proposed approach to generalize to various clutter distribution types (various $\nu$ values), and the ability of the proposed approach to suppress correlated clutter, since it succeeds in ``extracting'' embedded targets from the clutter.

\subsubsection{Number of Targets}\label{subsubsec:cg_ntargets}
\begin{figure}[!t]
\centering
    \includegraphics[width=0.4\textwidth]{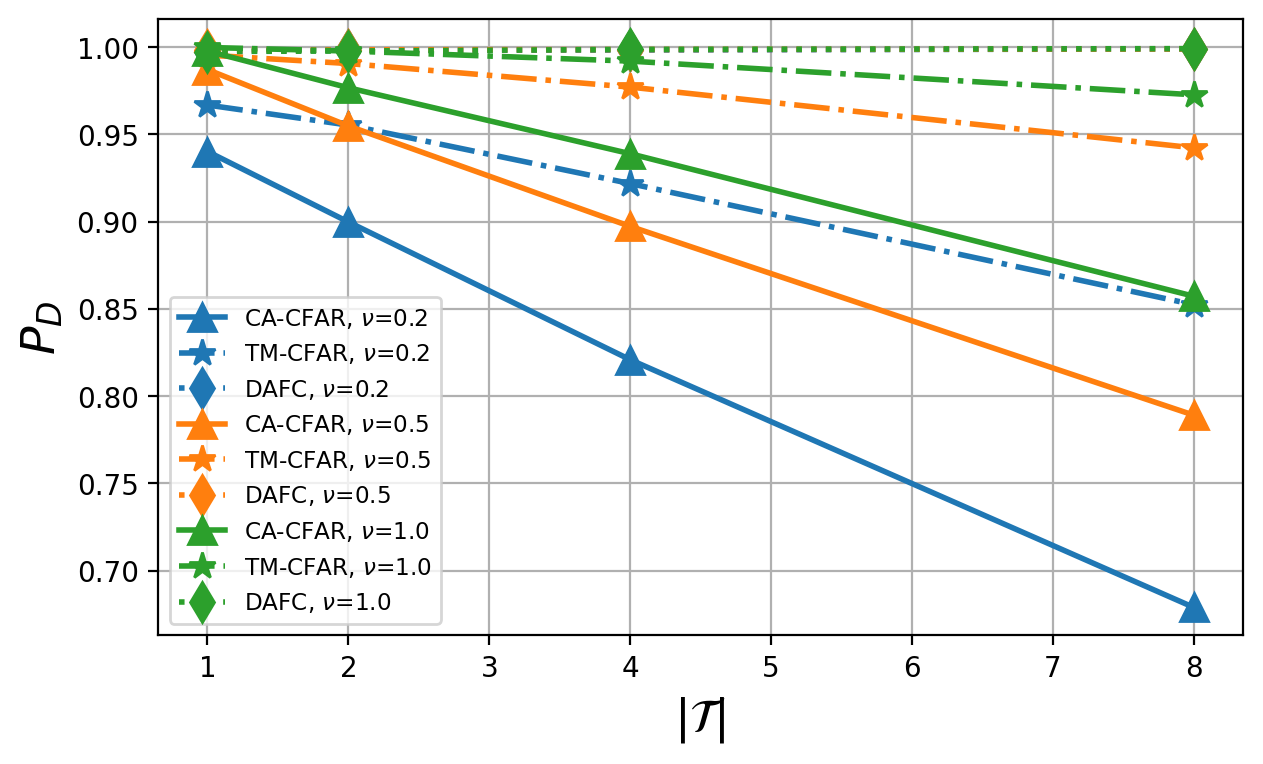}
    \caption{Detection performance with simulated clutter as a function of the number of targets with $\text{SCNR}=0 dB$ and $P_{FA}=5\cdot10^{-4}$ for various clutter ``spikeness'' $\nu\in\{0.2,\;0.5,\;1\}$.} \label{fig:CG_ntargets}
\end{figure}

The sensitivity of the proposed approach to the number of targets $\left|\mathcal{T}\right|$ is evaluated \textcolor{black}{in this experiment using clutter with targets as defined in~\ref{subsubsec:eval_sim_clutter_performance}}.
Fig.~\ref{fig:CG_ntargets} shows that the CA-CFAR \textcolor{black}{and \textcolor{black}{TM-CFAR}} performances degrade with decreasing $\nu$ and increasing the number of targets (due to the mutual target masking \cite{richards2010principles}).
On the other hand, The proposed approach shows robustness to clutter ``spikiness'' variation and demonstrates generalization capability to multiple targets.

\begin{figure}[!t]
\centering
\includegraphics[width=0.41\textwidth]{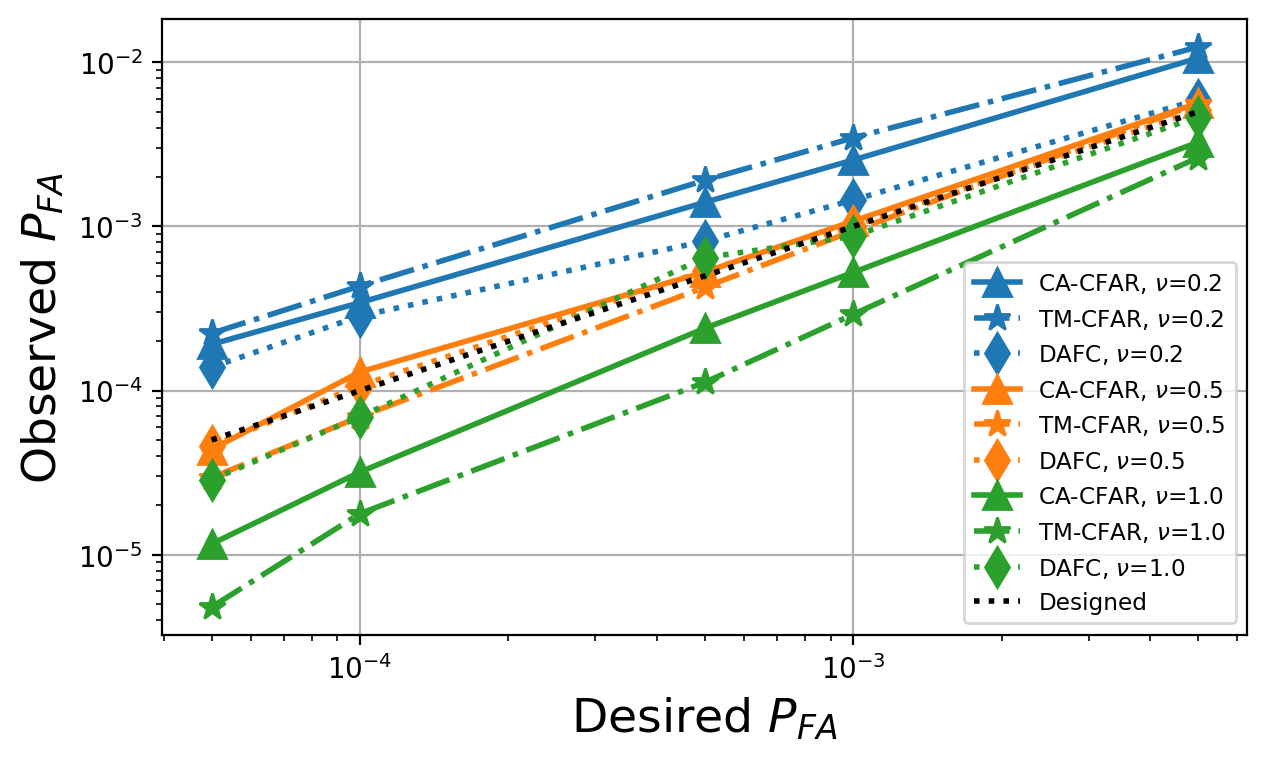}
\caption{\textcolor{black}{$P_{FA}$ verification for simulated clutter with ``spikeness'' levels $\nu\in\{0.2,\;0.5,\;1\}$. 
The desired $P_{FA}$ denotes the false-alarm probability obtained using the validation dataset, and the observed $P_{FA}$ denotes the false-alarm probability evaluated using the test dataset when the thresholds are according to the validation dataset.}}
\label{fig:CG_pfa_miss}
\end{figure}

\begin{figure*}[!t]
  \begin{subfigure}{0.42\textwidth}
    \caption{Clutter with targets, $P_{FA}=5 \cdot 10^{-4}$} \label{subfig:ipix_not_embedded_scnr}
    \includegraphics[width=\linewidth]{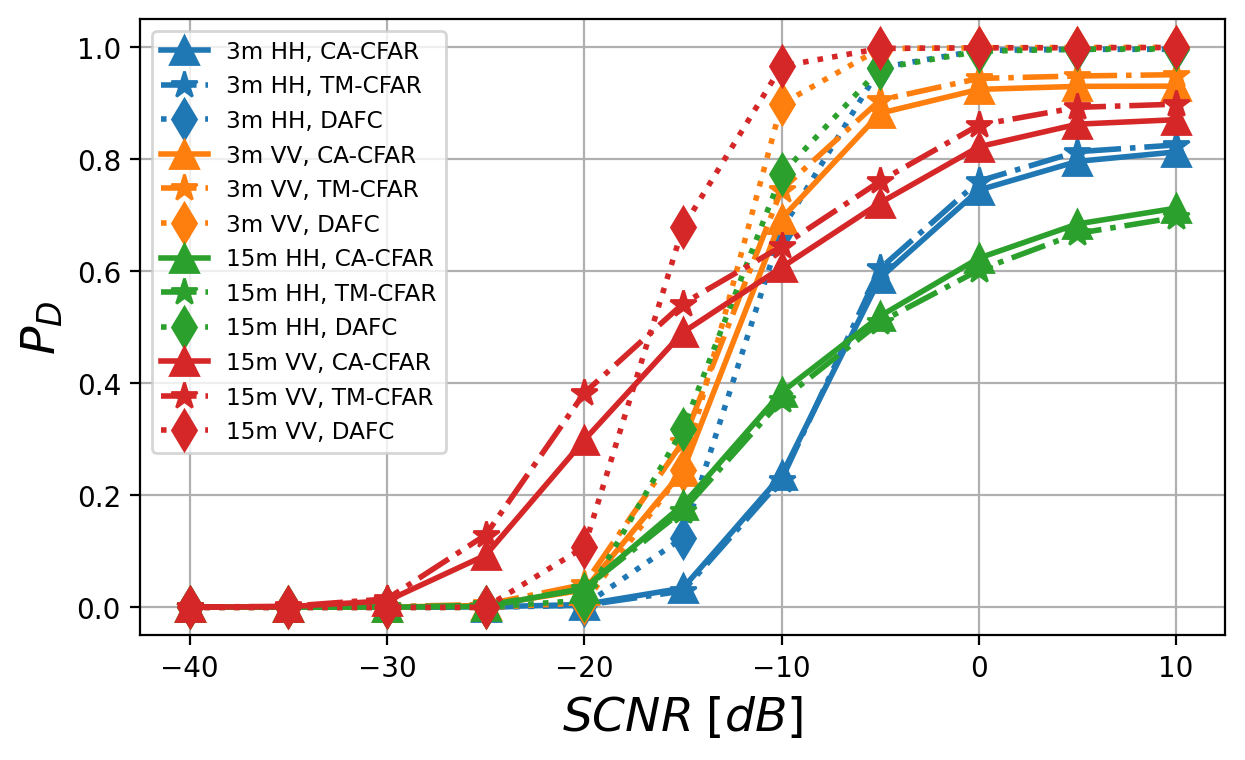}
  \end{subfigure}%
  \hspace*{\fill}   
  \begin{subfigure}{0.42\textwidth}
    \caption{Clutter with targets, $\text{SCNR}=0\;dB$} \label{subfig:ipix_not_embedded_pfa}
    \includegraphics[width=\linewidth]{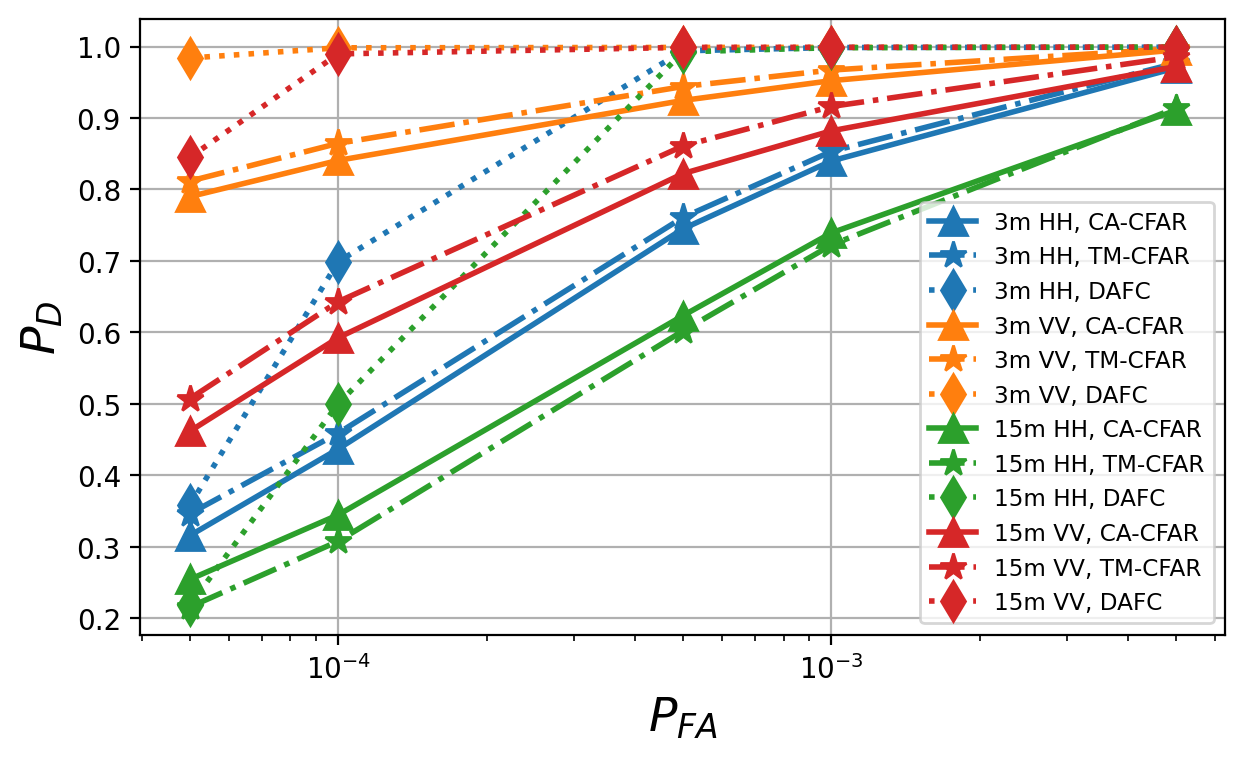}
  \end{subfigure}%
\caption{Detection results in scenarios with targets and recorded radar clutter for $3$m and $15$m range resolutions, and VV and HH polarizations. 
Subplot (a) considers constant $P_{FA}=5\cdot10^{-4}$ and subplot (b) considers SCNR $=0\;dB$.}
\label{fig:pd_ipix}
\end{figure*}

\begin{figure*}[!t]
  \begin{subfigure}{0.42\textwidth}
    \caption{Clutter with embedded targets, $P_{FA}=5 \cdot 10^{-4}$}
    \includegraphics[width=\linewidth]{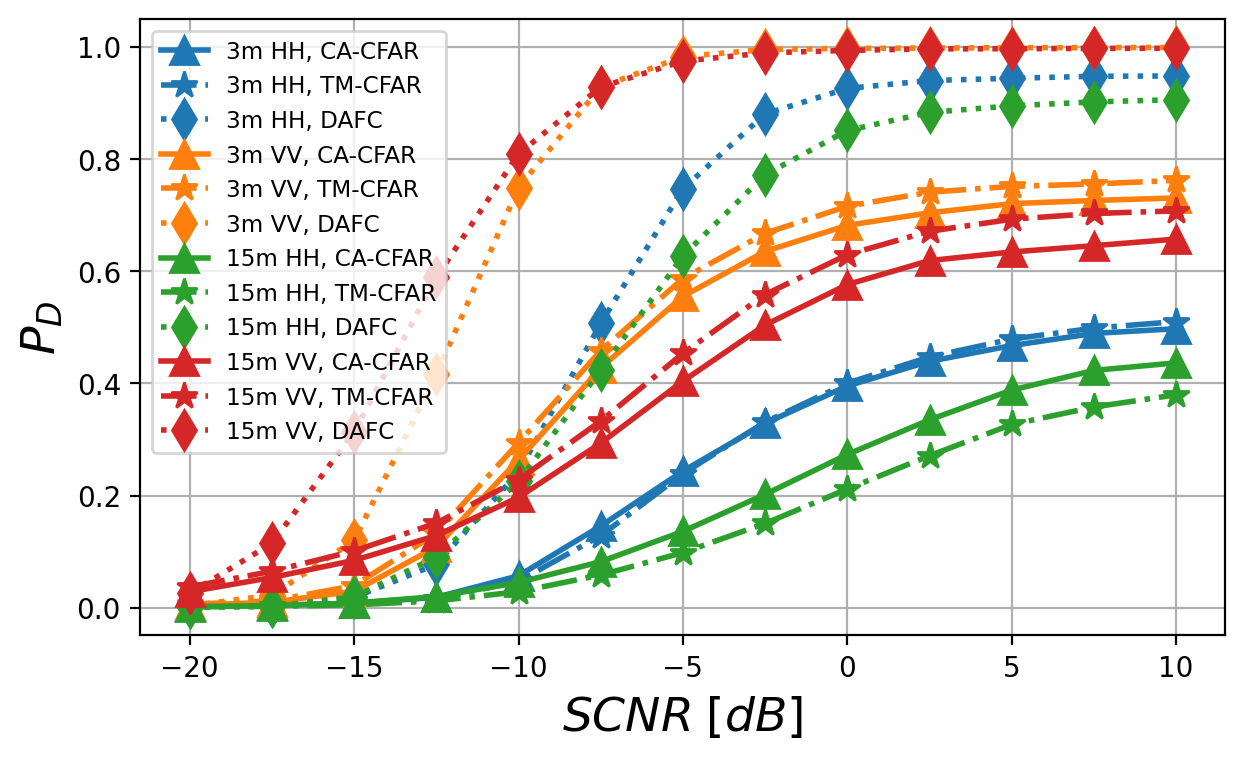}
    \label{subfig:ipix_embedded_scnr}
  \end{subfigure}%
  \hspace*{\fill}   
  \begin{subfigure}{0.42\textwidth}
    \caption{Clutter with embedded targets, $\text{SCNR}=0\;dB$}
    \includegraphics[width=\linewidth]{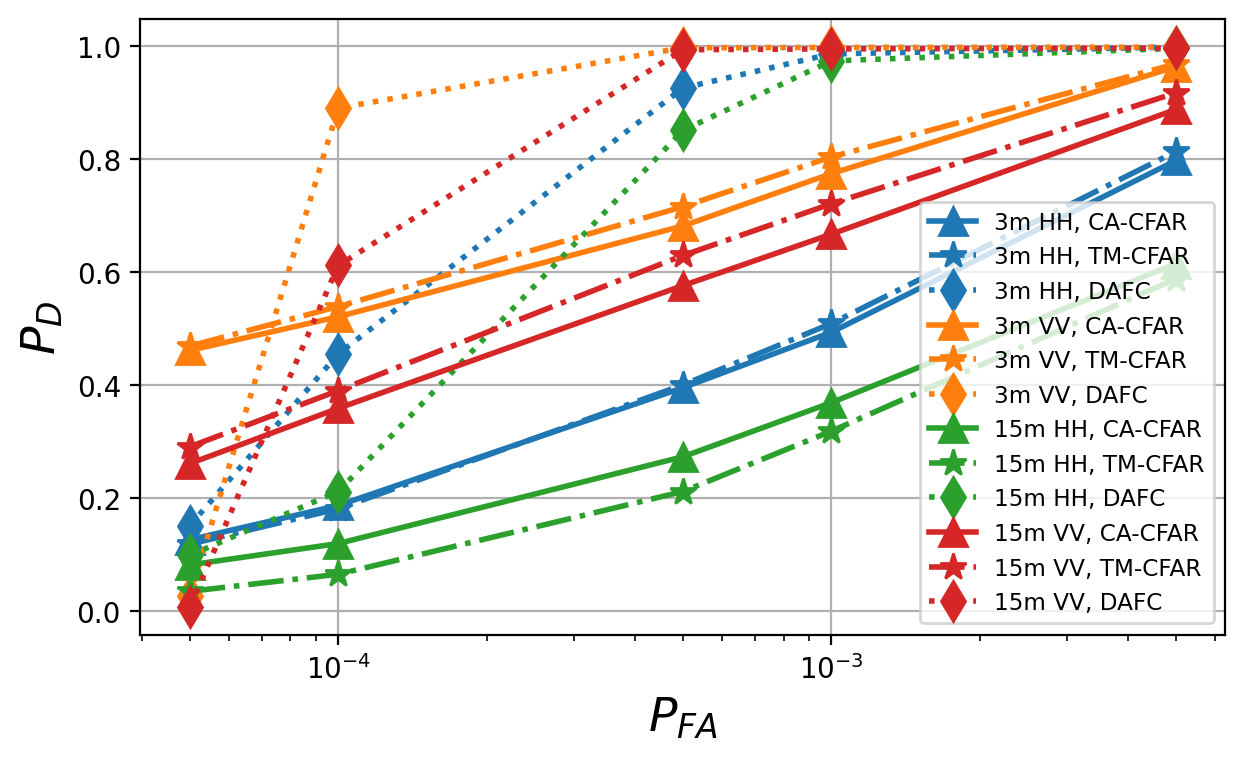}
    \label{subfig:ipix_embedded_pfa}
  \end{subfigure}%
\caption{Detection results in scenarios with embedded targets and recorded radar clutter for $3$m and $15$m range resolutions, and VV and HH polarizations. 
Subplot (a) considers constant $P_{FA}=5\cdot10^{-4}$ and subplot (b) considers SCNR $=0\;dB$.}
\label{fig:pd_ipix_embedded}
\end{figure*}

\subsubsection{\texorpdfstring{$P_{FA}$}{Lg}  Verification}\label{subsubsec:pfa_miss}

\textcolor{black}{The proposed approach is evaluated in terms of the detection threshold sensitivity in this subsection, via simulations of various types of clutter statistics determined by the parameter, $\nu$.
}


For each desired $P_{FA}$, the detection threshold for the proposed, the CA-CFAR \textcolor{black}{ and \textcolor{black}{TM-CFAR}} approaches, is selected using a validation dataset of $4,000$ frames with $\mathcal{T}\ne\emptyset$, and $2,000$ frames with $\mathcal{T}=\emptyset$. 
\textcolor{black}{For frames with $\mathcal{T}\ne\emptyset$, 4 targets and were considered with $\nu\sim\text{Unif}\left(0.1,1.5\right)$.}
The thresholds obtained according to this validation dataset are evaluated on test datasets with various clutter ``spikeness'' parameters, $\nu\in\{0.2,0.5,1\}$.

Fig.~\ref{fig:CG_pfa_miss} shows that the proposed approach is more robust than the \textcolor{black}{conventional CA-CFAR and \textcolor{black}{TM-CFAR} detectors}, in terms of variation in actual $P_{FA}$.
This can be explained by the fact that the clutter ``spikeness'' affects the adaptive detection threshold level in the CA-CFAR  and \textcolor{black}{\textcolor{black}{TM-CFAR}}~\cite{richards2010principles} detectors, and thus, their performances degrade in the $P_{FA}$ mismatch. 
In contrast, the NN models in the proposed approach learn a mapping (specifically a pmf) which is well generalized to variations in clutter statistics and therefore do not vary substantially when the clutter ``spikeness'' measure is changed.
\textcolor{black}{Although the CFAR property of the proposed approach is not guaranteed, Fig.~\ref{fig:CG_pfa_miss} shows that the false-alarm probability of the proposed approach is significantly less affected by the variation of $\nu$, compared to CA-CFAR and \textcolor{black}{TM-CFAR}.}

\subsection{Recorded Radar Clutter}\label{subsec:eval_real_world_clutter}
The performance of the proposed approach is evaluated in this Subsection using real recorded radar measurements. The radar clutter signals in each frame $\{\mathbf{c}_r\}_r$ are generated using recorded radar clutter from the McMaster IPIX database~\cite{IPIX}, as detailed in Subsection \ref{sec:data_gen}. Records that definitely do not contain any targets and contain only clutter-plus-noise echoes were selected. The following data files with HH and VV polarizations have been chosen for the $3m$ and $15m$ range resolution, respectively: $34, 36, 49, 52, 57, 86, 87, 88, 90, 98, 102, 103, 104, 105, 106$, $156, 165, 166$ and $35, 48, 55, 154$. The training and evaluation process was performed in a $k-fold$ cross-validation manner. The files with $3m$  range resolution were split into $6$ groups of $3$, and files with $15m$ range resolution were split into $4$ groups of $1$, where at each cross-validation iteration, a different group was left-out from training and used for test. 

We have found it beneficial to add a standardization step in the pre-processing flow described in Subsection~\ref{subsec:pre_process_flow}, which divides each element in $\mathbf{X}_1$ by the sample standard deviation computed over the element's column.
For the $15m$ range resolution files, the targets' ranges are sampled from the corresponding ranges and the chirp bandwidth $B$ is changed accordingly. A different pair of NN models for each range resolution and polarization were trained and evaluated on all considered SCNRs.
In addition, the clutter-plus-noise statistics is unknown, since this is real recorded radar data, hence the SCNR and CNR definitions in~\eqref{eq:scnr_def} are unavailable. Therefore, we resort to empirical methods to approximate the clutter-plus-noise energy within each sampled frame, by using the norm of the sampled real recorded data.

Figs.~\ref{fig:pd_ipix} and \ref{fig:pd_ipix_embedded} show the detection performance of the proposed approach to CA-CFAR and \textcolor{black}{\textcolor{black}{TM-CFAR}}, in a) clutter with targets and b) clutter with embedded targets, similarly to the experiment detailed in Subsection~\ref{subsubsec:eval_sim_clutter_performance}.
Note that all tested approaches achieve better performance for the VV polarization. This observation can be explained by the analysis of the polarization effects in the IPIX database~\cite{IPIX} in~\cite{1337463}, where it was shown that the HH amplitude is spikier than the VV amplitude~\cite{de2005cfar}.
Both Figs.~\ref{fig:pd_ipix} and \ref{fig:pd_ipix_embedded} show that the proposed approach outperforms the CA-CFAR and \textcolor{black}{\textcolor{black}{TM-CFAR detectors}} for the majority of tested SCNR values. 
\textcolor{black}{Similarly to Figs. \ref{fig:pd_sim_clutter} and \ref{fig:pd_sim_clutter_embedded}, it can be observed that  CA-CFAR and TM-CFAR do not reach $P_D=1$ for high SCNR, as a result of mutual target masking~\cite{richards2010principles}.
Fig.~\ref{subfig:ipix_embedded_scnr} shows a similar trend for the proposed approach in HH polarizations, but it significantly outperforms the CA-CFAR and TM-CFAR detectors.}


Furthermore, since these results are obtained from a cross-validation experiment on real data, it is a strong evidence for the generalization capability of the proposed approach to unseen data. 
Each file in the database contains clutter recordings from various dates, hours, azimuths, and range sections, hence the interference statistics contained in each file are different. Therefore, these results show that the proposed approach has an ability of \textcolor{black}{generalization to unseen data and clutter statistics.}

\subsection{Combination of NN-Based and Model-Based}\label{subsec:eval_combination_nn_sp}
\begin{figure}[!t]
\centering
\includegraphics[width=0.42\textwidth]{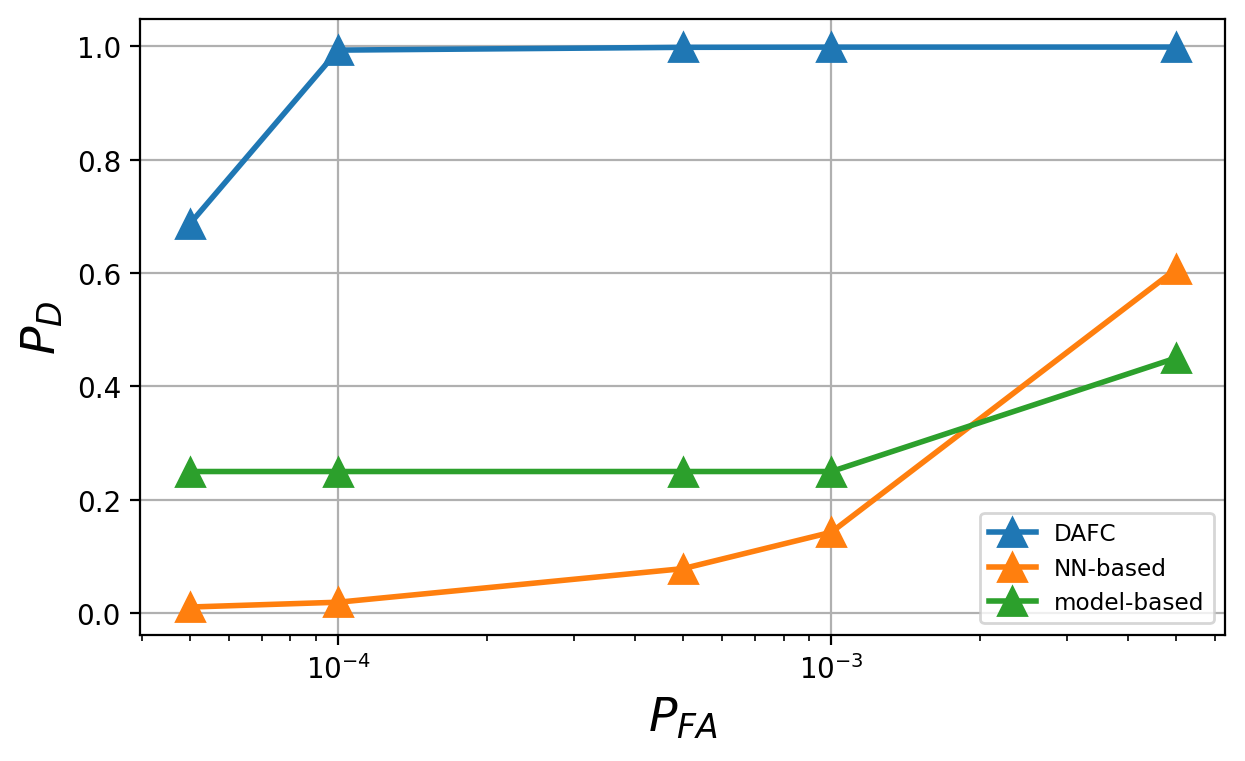}
\caption{ROC of various portions of the proposed algorithm with $\text{SCNR}=0\;dB$, and simulated clutter with $\nu=0.5$.}
\label{fig:detector_steps}
\end{figure}

The proposed approach combines NNs with model-based processing (projection to steering vector matrices). This Subsection demonstrates the significant performance gain provided by this combination.
Fig.~\ref{fig:detector_steps} shows the ROC of the proposed approach using simulated radar clutter with $\nu=0.2$.  Plots in Fig.~\ref{fig:detector_steps} show contributions of various components of the proposed approach. The blue line represents \textbf{Algorithm 1}, the orange line represents using only the NN-based performance in \textbf{Algorithm 1}, i.e. $\mathbf{Z}=\mathbf{1}_{d_R}\mathbf{1}^T_{d_V}$ (where $\mathbf{1}_d$ is a column vector of size $d$ whose entries are equal to one) without normalizing by $\text{max}$ in \textit{Step 3} and the green line represents detection using only projection-based signals in \textbf{Algorithm 1}: $\mathbf{y}_r=\mathbf{1}_{d_R},\mathbf{y}_v=\mathbf{1}_{d_R}$ which results in $\mathbf{U}=\mathbf{Z}$.

Notice that the NN-only-based detection suffers from inherent ambiguity since the operation $\mathbf{y}_r\mathbf{y}_v^T$ results in a Cartesian product of all possible range-Doppler combinations. On the other hand, projection-based processing (transform) does not mitigate the correlated clutter.
These results show the importance of combining the two approaches (NN- and model-based), since each of them alone performs poorly, whereas the combination delivers a substantial gain in the detection performance.

\subsection{Label Sparsity}
\begin{figure}[!t]
\centering
\includegraphics[width=0.42\textwidth]{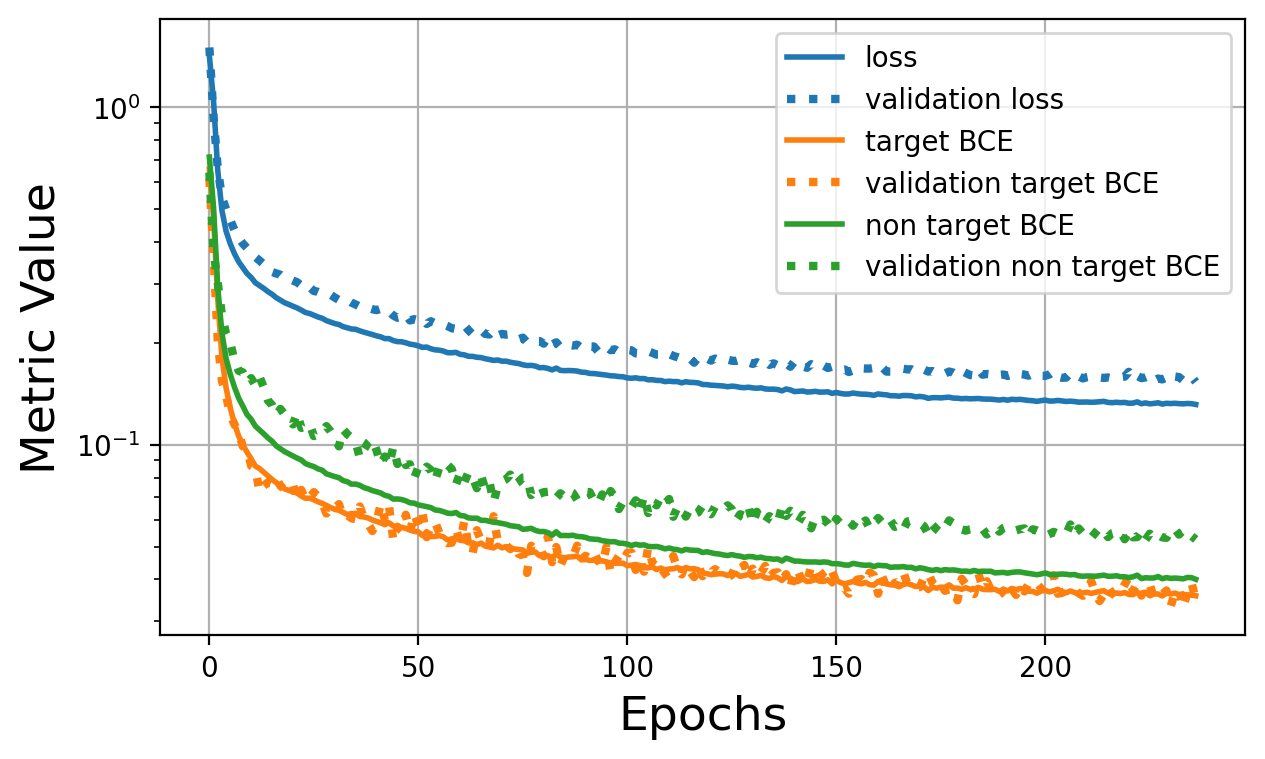}
\caption{Label sparsity effect investigation. Average learning curves of NNs training using recorded radar clutter echoes.}
\label{fig:BCE_fit}
\end{figure}

Evaluating the binary cross-entropy (BCE) criterion on target/non-target bins and loss function in~\eqref{eq:loss_function}, over the training and validation sets at each epoch can be used to evaluate the capabilities of the proposed approach.
Fig.~\ref{fig:BCE_fit} shows the average BCE and loss functions in range and Doppler domains, trained on the IPIX data and using the last $5 \%$ of each file ($3000$ pulses) as validation data.
The BCE for target-free bins converges, with lower values for training data. The BCE for target bins shows similar performance for validation and training sets. Fig.~\ref{fig:BCE_fit} clearly shows that the proposed NN architecture in combination with the class-balanced cross-entropy~\cite{cui2019class} generalizes to unseen data and overcomes the true labels sparsity challenge.

\subsection{Computational Complexity}

\textcolor{black}{
The computational complexity of the proposed approach is evaluated in this subsection considering $n=N=K=64$ and the range-Doppler map of size $N\times K$.
The computational complexity of the $N\times K$-sized 2D-FFT is the complexity of applying 1D-FFT per row followed by 1D-FFT per column, which is $\mathcal{O}\left(2n\cdot n\log n\right)$.
Considering $n=64$, the 2D-FFT complexity is $\mathcal{O}\left(768n\right)$.
The absolute squared value operator applied per range-Doppler bin,
with a complexity of $\mathcal{O}\left(n^2\right)$ adds additional complexity of $\mathcal{O}\left(64n\right)$.
The total complexity of the conventional range-Doppler transform is $\mathcal{O}\left(832n\right)$ for $n=64$.
For CA-CFAR, the additional complexity is associated with averaging the cells in the window, which results in the computational complexity of $\mathcal{O}\left(2n\right)$. 
For the \textcolor{black}{TM-CFAR}, the additional computation complexity is associated with sorting cells in the window. 
The \textcolor{black}{TM-CFAR} window contains $m=126$ cells, and the sorting operation takes $\mathcal{O}\left(m\log m\right)$ complexity, which is equivalent to $\mathcal{O}\left(13 n\right)$ for $n=64$.
Assuming parallel execution of CA-CFAR and \textcolor{black}{TM-CFAR} per range-Doppler bin, their complexity is  $\mathcal{O}\left(832n+2n\right)=\mathcal{O}\left(834n\right)$ and $\mathcal{O}\left(832n+13n\right)=\mathcal{O}\left(845n\right)$, respectively.
}

\textcolor{black}{
The proposed approach in \textbf{Algorithm 1} consists of $3$ steps. \textit{Step 1} contains NN feed-forward, and model-based projection, executed in parallel.
The DAFC operation consists of two consecutive matrix multiplications and applying activation functions.
Assuming parallel execution of matrix-vector multiplication, the complexity of $\mathbf{Z}\mathbf{w}$, where $\mathbf{Z}\in\mathbb{R}^{d_1\times d_2}$ and $\mathbf{w}\in\mathbb{R}^{d_2}$ is $\mathcal{O}\left(d_2\right)$. 
Namely, we assume that the dot product between the rows of $\mathbf{Z}$ and $\mathbf{w}$ is executed in parallel.
As detailed in Subsection~\ref{subsubsec:dafc_block}, the DAFC operation consists of two consecutive matrix multiplications. 
Consider $\mathbf{Z}\in\mathbb{R}^{d_1\times d_2}$ and $\mathbf{W}=\begin{bmatrix}\mathbf{w}_{d_1} & \dots & \mathbf{w}_{d_3}\end{bmatrix}\in\mathbb{R}^{d_1\times d_3}$. 
Then $\mathbf{ZW}=\begin{bmatrix}\mathbf{Zw}_{d_1} & \dots & \mathbf{Zw}_{d_3}\end{bmatrix}$ consists of $d_3$ matrix-vector multiplications of $\mathcal{O}\left(d_2\right)$ complexity, which is $\mathcal{O}\left(d_2d_3\right)$.
Therefore, the complexity of transforming an $H\times W$ input to an $H' \times W'$ output is $\mathcal{O}\left(WW' + HH'\right)$.
According to Table~\ref{tab:arch}, The complexity of the DAFC blocks is $\mathcal{O}\left(1024\cdot2n+128n\right)$ for the first block, $\mathcal{O}\left(256\cdot16n + 16 \cdot 2n \right)$ for the second block, and $\mathcal{O}\left(128\cdot4n + 4 \cdot 0.25n \right)$ for the third block.
The total complexity of the DAFC operations sums up to $\mathcal{O}\left(6817n\right)$.
The final FC layer's complexity is $\mathcal{O}\left(8n\right)$.
The complexity of bias addition activation functions was neglected in this analysis since these can also be executed in parallel.
Therefore, the computational complexity of a single NN instance consists of $\mathcal{O}\left(6825n\right)$.
}

\textcolor{black}{
Two NN instances can be executed in parallel, and the 2D-FFT plus absolute operation exceeds a single NN feed-forward computational complexity. Therefore, the computational complexity of \textit{Step 1} in \textbf{Algorithm 1} is $\mathcal{O}\left(6825n\right)$.
The computational complexity of \textit{Step 2} and \textit{Step 3} is negligible, since it contains per range-Doppler bin operations, which can be executed in parallel.
In addition, the pre-processing complexity, $\mathcal{O}\left(n\right)$, is relatively low and can be executed during the following frame. Thus, the total computational complexity of the proposed approach is $\mathcal{O}\left(6825n\right)$. 
}

\textcolor{black}{
The computational complexities of the proposed approach, the CA-CFAR, and the \textcolor{black}{TM-CFAR} are summarized in Table~\ref{tab:complexity}.
\begin{table}[ht]
\begin{center}
 \begin{tabular}{m{2.5cm} m{2.5cm} m{2cm}}
 \hline
 CA-CFAR & \textcolor{black}{TM-CFAR} & DAFC\\ [1.0ex] 
 \hline\hline
 $\mathcal{O}\left(834n\right)$ & $\mathcal{O}\left(845n\right)$ & $\mathcal{O}\left(6825n\right)$ \\ [1.0ex]
 \hline
\end{tabular}
\end{center}
\caption{\label{tab:complexity} Computational complexity in terms of $n=N=K=64$.}
\end{table}
Although the computational complexity of the proposed approach is $8$ times higher than the complexity of the CA-CFAR and  \textcolor{black}{TM-CFAR} detectors, the proposed NN architecture consists of consecutively performed matrix multiplications and element-wise activation functions. 
Therefore, the proposed method's computational complexity can be significantly reduced using an appropriate implementation of basic linear algebra operations, implemented using hardware accelerators such as graphics processing unit (GPU) or digital signal processor (DSP).
}

\section{CONCLUSION}\label{sec:conclusion}
This work addressed the problem of multiple target detection in the range-Doppler domain in the presence of correlated heavy-tailed clutter. 
A NN-based approach was proposed to learn a complex nonlinear transform for clutter mitigation in LFM radar, and its superiority over the conventional CA-CFAR, \textcolor{black}{\textcolor{black}{TM-CFAR}, and ANMF} detectors was demonstrated.
A novel DAFC processing block was introduced to utilize the structure of information encoded in each complex \textit{fast-time} $\times$ \textit{slow-time} radar echo and to transform it using a sparse set of parameters.
A unified NN-based architecture incorporating the DAFC block was proposed for the problem of multiple target detection in range or Doppler domains, separately. 
The generalization capability of the unified DAFC-based NNs to various SCNRs and clutter conditions was demonstrated via multiple tests. This generalization capability simplifying the NN training and reduces implementation complexity.

The proposed architecture was used to design a range-Doppler detector that uses the NNs outputs as a pmf to re-weight a conventional projection into steering vectors.
The performance of the proposed approach was evaluated using simulated correlated heavy-tailed clutter and a database of recorded heavy-tailed radar clutter echoes.
\textcolor{black}{The superiority of the proposed approach over the ANMF, the conventional CA-CFAR and the robust \textcolor{black}{TM-CFAR}} detectors was demonstrated in multiple tested scenarios. 
Robustness to increasing the number of targets was observed for the AWGN and correlated heavy-tailed clutter cases. 
A significant performance advantage was demonstrated for various clutter ``spikiness'' conditions in terms of probability of detection and detection threshold sensitivity.
Real data-based experiments demonstrated the strong generalization capabilities of the proposed approach to unseen data, containing various clutter statistics.

\bibliographystyle{IEEEtran}
\bibliography{main}

\begin{IEEEbiography}[{\includegraphics[width=1in,height=1.25in,clip,keepaspectratio]{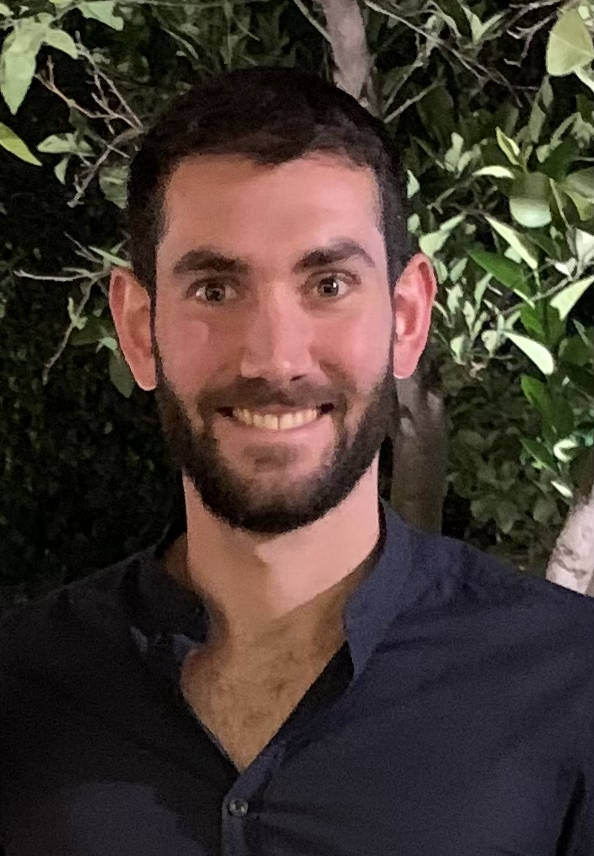}}]%
{Stefan Feintuch} received B.sc. and M.Sc. in electrical and computer engineering from the Ben-Gurion University of the Negev, Beer Sheva, Israel, in 2022 and 2023, respectively.
His research interests include machine learning, deep learning, radars, and signal processing.
\end{IEEEbiography}

\begin{IEEEbiography}[{\includegraphics[width=1in,height=1.25in,clip,keepaspectratio]{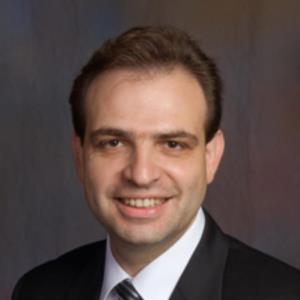}}]%
{Igal Bilik}(S'03-M'06-SM'21) received B.Sc., M.Sc., and Ph.D. degrees in electrical and computer engineering from the Ben-Gurion University of the Negev, Beer Sheva, Israel, in 1997, 2003, and 2006, respectively. During 2006–2008, he was a postdoctoral research associate in the Department of Electrical and Computer Engineering at Duke University, Durham, NC. During 2008-2011, he has been an Assistant Professor in the Department of Electrical and Computer Engineering at the University of Massachusetts, Dartmouth. During 2011-2019, he was a Staff Researcher at GM Advanced Technical Center, Israel, leading automotive radar technology development. Between 2019-2020 he was leading Smart Sensing and Vision Group at GM R\&D, where he was responsible on development state-of-art automotive radar, lidar and computer vision technologies. Since Oct. 2020, Dr. Bilik is an Assistant Professor in the School of Electrical and Computer Engineering at the Ben-Gurion University of the Negev. Since 2020, he is a member of IEEE AESS Radar Systems Panel and a vice-Chair of Civilian Radar Committee. Dr. Bilik is an Acting Officer of IEEE Vehicular Technology Chapter, Israel. Dr. Bilik  has more than 170 patent inventions, authored more than 60 peer-reviewed academic publications, received the Best Student Paper Awards at IEEE RADAR 2005 and IEEE RADAR 2006 Conferences, Student Paper Award in the 2006 IEEE 24th Convention of Electrical and Electronics Engineers in Israel, and the GM Product Excellence Recognition in 2017.
\end{IEEEbiography}

\begin{IEEEbiography}[{\includegraphics[width=1in,height=1.25in,clip,keepaspectratio]{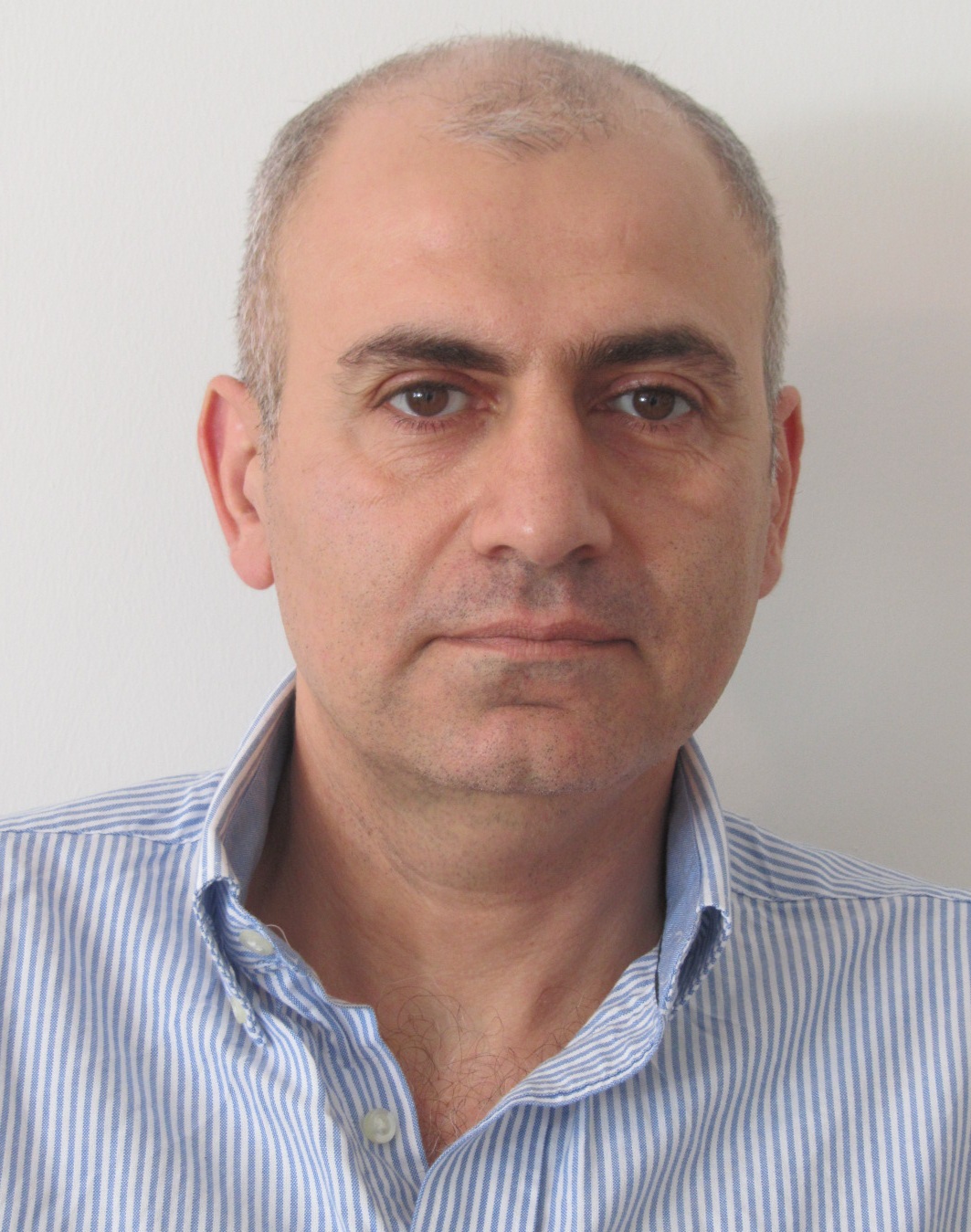}}]%
{Joseph Tabrikian} (Fellow, IEEE) received the B.Sc., M.Sc., and Ph.D. degrees in Electrical Engineering from the Tel-Aviv University, Tel-Aviv, Israel, in 1986, 1992, and 1997, respectively. During 1996–1998 he was with the Department of Electrical and Computer Engineering (ECE), Duke University, Durham, NC as an Assistant Research Professor. In 1998, he joined the Department of ECE, Ben-Gurion University of the Negev, Beer-Sheva, Israel, and served as the department head during 2017-2019. In May 2019 he established the school of ECE and served as its head till August 2021. He served as an Associate Editor (AE) for the IEEE Transactions on Signal Processing during 2001–2004 and 2011-2015, and is currently a Senior Area Editor (SAE) for these transactions. He served as AE and SAE of the IEEE Signal Processing Letters during 2012-2015 and 2015-2018, respectively. He was a member of the IEEE Sensor Array and Multichannel (SAM) technical committee during 2010-2015 and was the technical program co-chair of the IEEE SAM 2010 workshop. During 2015-2021 he served as a member of the Signal Processing for Multisensor Systems (SPMuS) of EURASIP and during 2017-2022 he was a member IEEE SPTM technical committee. He is co-author of 7 award-winning papers in IEEE conferences and workshops. His research interests include estimation and detection theory, learning algorithms, and radar signal processing. 
\end{IEEEbiography}

\begin{IEEEbiography}
[{\includegraphics[width=1in,height=1.25in,clip,keepaspectratio]{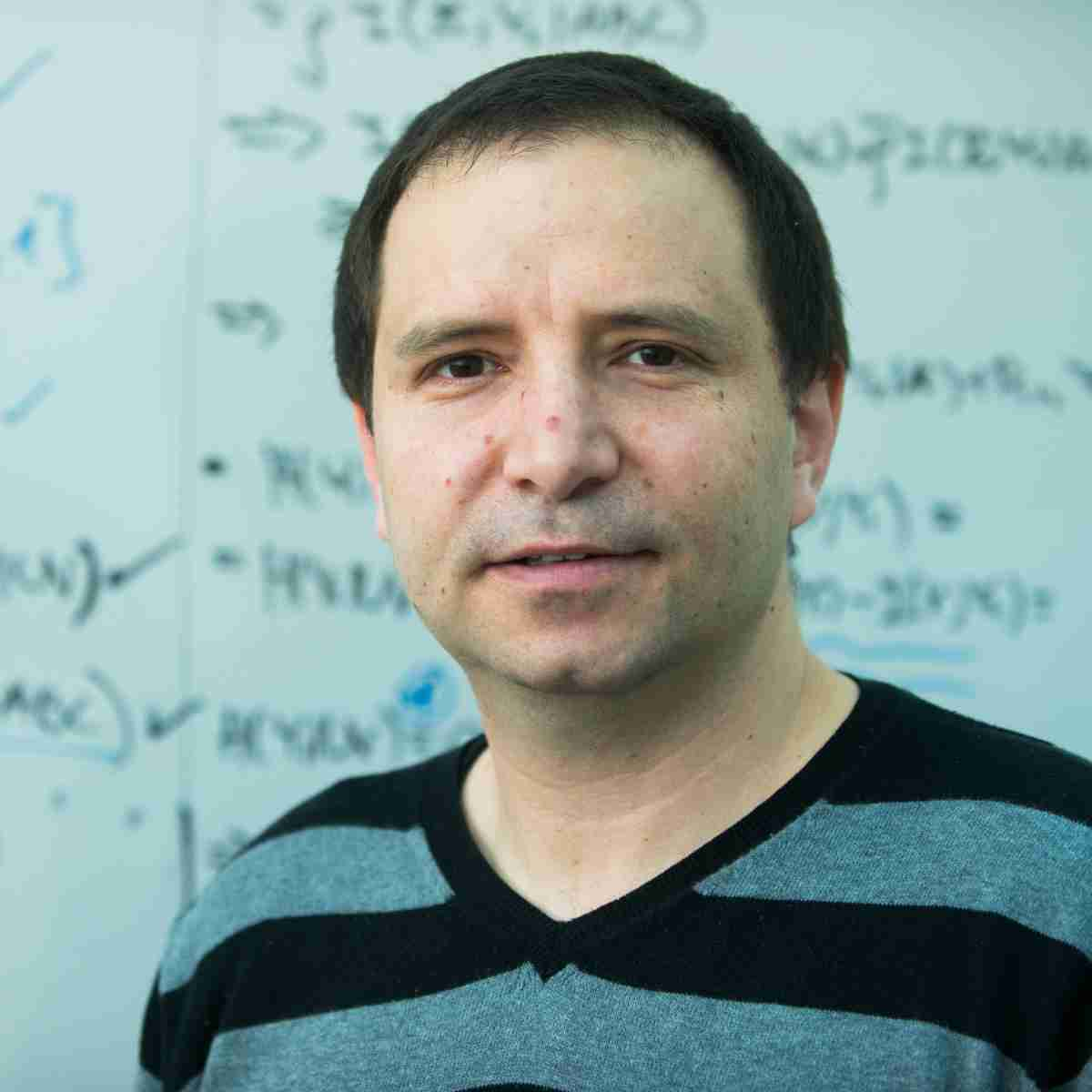}}]%
{Haim H. Permuter} (M'08-SM'13) received his B.Sc.\@ (summa cum laude) and M.Sc.\@(summa cum laude) degrees in Electrical and Computer Engineering from the Ben-Gurion University, Israel, in 1997 and 2003, respectively, and the Ph.D. degree in Electrical  Engineering from Stanford University, California in 2008. Between 1997 and 2004, he was an officer at a research and development unit of the Israeli Defense Forces. Since 2009 he is with the department of Electrical and Computer Engineering at Ben-Gurion University where he is currently  a professor, Luck-Hille Chair in Electrical Engineering. Haim also serves as head of the communication,cyber/ and information track in his department. Prof. Permuter is a recipient of several awards, among them the Fullbright Fellowship, the Stanford Graduate Fellowship (SGF), Allon Fellowship, and and the U.S.-Israel Binational Science Foundation Bergmann Memorial Award. Haim served on the editorial boards of the IEEE Transactions on Information Theory in 2013-2016 and has been reappointed again in 2023.
\end{IEEEbiography}

\end{document}